 \documentclass[final,onecolumn,3p,times,sort&compress,12pt]{elsarticle}

\usepackage{graphicx}% Include figure files
\usepackage{dcolumn}% Align table columns on decimal point
\usepackage{bm}% bold math
\usepackage{epstopdf}% pdflatex command active even for eps figures
\usepackage{bm,hyperref,color,breakurl}

	\usepackage{fancyheadings}
\renewcommand{\thispagestyle}[1]{} % do nothing

\begin{document}

%\preprint{}

\title{Thermodynamic model of a solid with RKKY interaction and magnetoelastic coupling}


\author[a1]{T. Balcerzak}
\ead{t\_balcerzak@uni.lodz.pl}
%\thanks{}
\author[a1]{K. Sza\l{}owski\corref{cor1}}
\ead{kszalowski@uni.lodz.pl}
%\thanks{}
\author[a3]{M. Ja\v{s}\v{c}ur}
%\thanks{}
\address[a1]{Department of Solid State Physics, Faculty of Physics and Applied Informatics,\\
University of \L\'{o}d\'{z}, ulica Pomorska 149/153, 90-236 \L\'{o}d\'{z}, Poland}

\address[a3]{Department of Theoretical Physics and Astrophysics, Faculty of Science, \\P. J. \v{S}\'{a}f\'{a}rik University, Park Angelinum 9, 041 54 Ko\v{s}ice, Slovak Republic}

\cortext[cor1]{Corresponding author}


\date{\today}

\begin{abstract}
Thermodynamic description of a model system with magnetoelastic coupling is presented. The elastic, vibrational, electronic and magnetic energy contributions are taken into account. The long-range RKKY interaction is considered together with the nearest-neighbour direct exchange. The generalized Gibbs potential and the set of equations of state are derived, from which all thermodynamic functions are self-consistently obtained. Thermodynamic properties are calculated numerically for FCC structure for arbitrary external pressure, magnetic field and temperature, and widely discussed. In particular, for some parameters of interaction potential and electron concentration corresponding to antiferromagnetic phase, the existence of negative thermal expansion coefficient is predicted.  
\end{abstract}


\maketitle

%section 1
\section{Introduction}
\label{sec1}
Thermodynamics of solids with magnetoelastic couplings is a subject of extensive interest of solid state physicists since many years in its various aspects \cite{salinas_one-dimensional_1973, bergman_critical_1976, chakrabarti_critical_1977, chakrabarti_critical_1979, henriques_effective_1987, boubcheur_effect_1999, massimino_effect_2000, boubcheur_effects_2001, cannavacciuolo_critical_2005, li_magnetoelastic_2010, chakrabarti_critical_1980, chakrabarti_critical_1982, chakrabarti_critical_1980-1, pytte_spin-phonon_1965, barma_phonon-induced_1975, ngo_monte_2002, liarte_compressible_2009, sobkowicz_ising_1996, vecchini_magnetoelastic_2010, zorko_magnetic_2015, bergman_exactly_1973, strecka_spontaneous_2012, strecka_spin-phonon_2012, rutkowski_pressure_2010, leger_pressure_1972, gehring_pressure-induced_2008, radomska_magnetic_2001,  amaral_magnetoelastic_2004, matsumoto_ehrenfest_2005, alho_influence_2010, szuszkiewicz_spin-wave_2006,bean_magnetic_1962,callen_static_1963,callen_magnetostriction_1965,brown_theory_1965,alben_magnetoelastic_1969,del_moral_magnetostriction_2007,lu_general_2015, buchelnikov_magnetoelastic_2002,balcerzak_self-consistent_2017,szalowski_thermodynamics_2018}. The magnetoelastic interactions are responsible for such effects as the magnetostriction  \cite{callen_static_1963,callen_magnetostriction_1965,alben_magnetoelastic_1969} and piezomagnetism, which are important from the point of view of application. As another direct consequence of the presence of magnetoelastic coupling, one can mention the pressure influence on the magnetic phase transition temperature, which has been discussed in numerous works \cite{bergman_critical_1976, liarte_compressible_2009, rutkowski_pressure_2010, leger_pressure_1972, gehring_pressure-induced_2008, radomska_magnetic_2001,  bean_magnetic_1962, santana_anomalous_2011, de_oliveira_magnetocaloric_2014}. The studies involve both model systems and specific materials, among which a particularly important class of magnetic semiconductors can be mentioned \cite{ gonzalez_szwacki_gamnas_2015,lim_exchange_2011,sollinger_exchange_2010,gryglas-borysiewicz_hydrostatic_2010,lim_effect_2009,csontos_pressure-induced_2005,csontos_effect_2004,suski_ferromagnetism_1987}. Moreover, the contemporarily studied magneto-caloric materials also essentially rely on the existence of the coupling between the crystalline lattice and the magnetic subsystem \cite{amaral_magnetoelastic_2004,alho_influence_2010,santana_anomalous_2011, de_oliveira_magnetocaloric_2014, piazzi_magnetocaloric_2015}, which influences the vital parameters of these materials.

In a common approach, the magnetoelastic coupling arises from the fact that the magnetic exchange integral between magnetic moments depends on their mutual distance \cite{salinas_one-dimensional_1973,henriques_effective_1987,boubcheur_effect_1999, massimino_effect_2000, boubcheur_effects_2001,chakrabarti_critical_1980-1,barma_phonon-induced_1975, ngo_monte_2002,bergman_exactly_1973, strecka_spontaneous_2012,radomska_magnetic_2001,sollinger_exchange_2010,szuszkiewicz_spin-wave_2006,balcerzak_self-consistent_2017, szalowski_thermodynamics_2018, iwashita_curie_1984}, which makes the magnetic energy volume dependent. On the other hand, the volume is an indispensable parameter occurring in other, non-magnetic, parts of the total energy, as for instance, the elastic potential energy, vibrational energy, as well as the electronic one. 

For the system in stable equilibrium, the total energy must take the minimum value. This can be achieved when the volume and magnetization of the system are treated as variational parameters, whereas the external pressure, magnetic field and temperature are independent and fixed variables. The variational approach leads to the set of equations of state in which the volume and magnetization are interrelated and determined by the rest of independent variables. Thus, the influence of the external pressure on the magnetic variational parameter (magnetization) can be manifested, in addition to the expected change of the volume. On the other hand, the external magnetic field influences, via magnetic energy, the volume of the system, in addition to the expected change of the magnetization.

In our previous papers \cite{balcerzak_self-consistent_2017, szalowski_thermodynamics_2018}, the thermodynamic model for the magnetoelastic couplings was presented, for the simplest case when the magnetic interaction between localized spins was of Heisenberg type. The energy of itinerant electrons was not considered in that approach, thus restricting the model to magnetic insulators. However, the energy of electron subsystem is important in such systems as metals, being responsible for the metallic bonds and contributing to the elastic properties. On the other hand, the presence of electron gas enables the long-range Ruderman-Kittel-Kasuya-Yosida (RKKY) indirect interaction between localized spins \cite{ruderman_indirect_1954, kasuya_theory_1956,yosida_magnetic_1957}. The exchange interaction in RKKY model is oscillating vs. distance, and its amplitude is volume dependent. Thus, in a natural way it is sensitive to the volume deformation.

Since up to now studies of magnetoelastic properties with RKKY interaction included seem to be rather unexploited area, the aim of the present paper is to fill the existing gap. We will make use of the underlying methodology developed in our previous paper \cite{balcerzak_self-consistent_2017}, and extend the approach by taking into account the itinerant electron energy in Hartree-Fock approximation. Then, the long-range RKKY interaction will be included in addition to the nearest-neighbour (NN) direct Heisenberg interaction. Thus, in the present model the magnetoelastic couplings have two sources: the volume dependence of the NN Heisenberg exchange integral, as well as the long-range RKKY interaction. In addition, when the external magnetic field is present, the effective gyromagnetic factor in the RKKY Hamiltonian occurs to be volume dependent \cite{balcerzak_rkky_2006}. In our opinion, all thses features make the present model interesting enough and much more complete than in the previous approach \cite{balcerzak_self-consistent_2017}, since all essential energy contributions to the total Gibbs energy are now taken into account.

The paper is organized as follows: in the next, theoretical, section the formalism will be presented. It contains a self-consistent thermodynamic methodology developed for the complex systems with many variables, including derivation of the generalized Gibbs potential and the equations of state. Some complementary formulas are placed in the Appendix. In the third section the exemplary numerical results will be presented in figures and discussed. They concern calculation of various thermodynamic parameters in the presence of magnetoelastic coupling. The calculations are performed for a model FCC lattice with NN and RKKY interaction. A comparison of the results for different electron concentrations, which correspond to the existence of ferromagnetic or antiferromagnetic phases, is made there. In the last section, the paper will be summarized and the conclusions will be drawn.

%section 2
\section{Theoretical model}
\label{sec2}

The Gibbs free energy of a system is assumed in the form of:\\
\begin{equation}
\label{eq1}
G=F_{\varepsilon}+F_{\rm D}+F_{\rm el}+pV+G_{m}
\end{equation}
where $F_{\varepsilon}$ is the elastic (static) Helmholtz energy, $F_{\rm D}$ is the vibrational (thermal) Helmholtz energy in the Debye approximation, $F_{\rm el}$ is the electronic Helmholtz energy in the Hartree-Fock approximation, $p$ is the external pressure, $V$-volume of the system, and $G_{m}$ is the Gibbs energy of magnetic subsystem with RKKY interaction. These energy components will be presented below.

\subsection{The elastic (static) subsystem}

The elastic energy $F_{\varepsilon}$ can be found on the basis of the Morse potential \cite{morse_diatomic_1929,girifalco_application_1959,lincoln_morse-potential_1967}. Considering the atomic pairs, where one atom stays in the centre of the system of coordinates and the second atom is situated on the $k$-th coordination sphere of radius $r_{k}$, the potential energy is given by: 
\begin{equation}
\label{eq2}
U(r_{k})=D\left(1-e^{-\alpha \left(r_{k}-r_0\right)/r_0}\right)^2.
\end{equation}
The pair-potential contains three fitting parameters: potential depth $D$, dimensionless asymmetry parameter $\alpha$ and the distance $r_0$ where the potential has its minimum.\\
We will assume that for the crystals with cubic symmetry the radius of $k$-th coordination sphere, $r_{k}$, can be expressed in terms of the isotropic volume deformation $\varepsilon$, namely:
\begin{equation}
\label{eq3}
r_{k}=r_{k,0}\left(1+\varepsilon\right)^{1/3}
\end{equation}
where $r_{k,0}$ is the radius of a non-deformed sphere and the isotropic volume deformation $\varepsilon$ is defined by the equation:
\begin{equation}
\label{eq4}
V=V_0\left(1+\varepsilon\right)=\frac{N}{z_0}a_{0}^{3}\left(1+\varepsilon\right).
\end{equation}
In Eq.(\ref{eq4}), $V_0 = V(p=0, H^{z}=0, T=0)$ is the volume of a non-deformed system (NDS) for $\varepsilon=0$, which is assumed at pressure $p=0$, magnetic field $H^{z}=0$ and temperature $T=0$. In the same formula, $N$ is the number of atomic sites, $z_0$ stands for the number of atoms per elementary cell, and the lattice constant of a non-deformed cubic cell is denoted by $a_0$. \\

It is convenient to use the pair-potential energy after shifting it by a constant value, $U(r_{k,0})$, in order to set zero Helmholtz energy $F_{\varepsilon}(\varepsilon=0)=0$ for a non-deformed crystal. The total elastic energy can be written as a sum over all the interacting pairs. For isotropic system the sum can be conveniently performed over the coordination zones with radii $r_{k,0}$ and the coordination numbers $z_k$. Finally, the elastic energy can be presented in the form of \cite{balcerzak_self-consistent_2017}
\begin{equation}
\label{eq5}
F_{\varepsilon}=\frac{N}{2}D\sum_{k}^{}{z_k\left\{ \left[1-e^{-\alpha \left(x\frac{r_{1,0}}{r_0}\frac{r_{k,0}}{a_{0}}\left(1+\varepsilon \right)^{1/3}-1\right)}\right]^2 - 
\left[1-e^{-\alpha \left(x\frac{r_{1,0}}{r_0}\frac{r_{k,0}}{a_{0}}-1\right)}\right]^2 \right\}}
\end{equation}
where the nearest neighbour normalized distance, $r_{1,0}/r_{0}$, can be found from the minimum conditions for the total Gibbs energy of a non-deformed crystal, whereas $r_{k,0}/a_{0}$ ratios and the coordination numbers $z_k$ are characteristic of a given crystallographic structure and can be found numerically. We have also introduced the coefficient $x$ relating the lattice constant and NN distance of a non-deformed lattice, namely $a_0=xr_{1,0}$. This coefficient is characteristic of a given lattice. For instance, for FCC structure $x=\sqrt{2}$, whereas $z_0=4$.  
Thus, the expression (\ref{eq5}) presents elastic energy for arbitrary isotropic deformation $\varepsilon$ with the assumption that $F_{\varepsilon}(\varepsilon=0)=0$.\\
The change of the elastic energy vs. volume is a source of elastic pressure:
\begin{eqnarray}
\label{eq6}
p_{\varepsilon}&=&-\left(\frac{\partial F_{\varepsilon}}{\partial V}\right)_T=-\frac{1}{V_0}\left(\frac{\partial F_{\varepsilon}}{\partial \varepsilon}\right)_T=\nonumber\\
&=&-\frac{x}{3}\frac{N}{V_0}D\alpha \frac{r_{1,0}}{r_0}\sum_{k=1}^{}{z_k \frac{r_{k,0}}{a_{0}}\left[1-e^{-\alpha \left(x\frac{r_{1,0}}{r_0}\frac{r_{k,0}}{a_{0}}\left(1+\varepsilon \right)^{1/3}-1\right)}\right]
\frac{e^{-\alpha \left(x\frac{r_{1,0}}{r_0}\frac{r_{k,0}}{a_{0}}\left(1+\varepsilon \right)^{1/3}-1\right)}}{\left(1+\varepsilon \right)^{2/3}}}.
\end{eqnarray}
This pressure should be taken into account together with other pressure contributions keeping the system in equilibrium.

\subsection{The vibrational (Debye) subsystem}

The vibrational energy is taken in the Debye approximation and for the arbitrary temperature $T$ can be presented as \cite{wallace_thermodynamics_1972} \\
\begin{equation}
\label{eq7}
F_{\rm D}=N\left[\frac{9}{8}k_{\rm B}T_{\rm D}+3k_{\rm B}T\ln \left(1-e^{-y_{\rm D}}\right)-3k_{\rm B}T\frac{1}{y_{\rm D}^3}\,\int_{0}^{y_{\rm D}}{\frac{y^3}{e^{y}-1}dy}\right],
\end{equation}
where $y_{\rm D}=T_{\rm D}/T$ and $T_{\rm D}$ is the Debye temperature. 

The Debye temperature is volume-dependent and can be expressed in the approximate form \cite{matsui_high_2010}
\begin{equation}
\label{eq8}
T_{\rm D}=T_{\rm D}^0
e^{\gamma_{\rm D}^0\left[1-\left(1+\varepsilon \right)^q\right]/q}.
\end{equation}
In Eq.(\ref{eq8}) $T_{\rm D}^0$ and $\gamma_{\rm D}^0$ are the Debye temperature and Gr\"uneisen parameter \cite{gruneisen_theorie_1912}, respectively, which are taken at $T=0$, $p=0$ and $H^{z}=0$, i.e., for non-deformed system, whereas $q$ is a constant. It has been shown that in the case of Morse potential in 3D systems the Gr\"uneisen parameter $\gamma_{\rm D}^0$ can be expressed as
\cite{krivtsov_derivation_2011}:
\begin{equation}
\label{eq9}
 \gamma_{\rm D}^0=\left(3\alpha-2\right)/6,
 \end{equation}
which, via elastic potential parameter $\alpha$, introduces anharmonicity to the Debye model.\\
The Debye integral in Eq.(\ref{eq7}) can be calculated from the exact formula \cite{dubinov_exact_2008,wood_computation_1992,balcerzak_self-consistent_2017}:
\begin{eqnarray}
\label{eq10}
\int_{0}^{y_{\rm D}}{\frac{y^3}{e^{y}-1}\,dy}&=&
\frac{1}{15}\pi^4-3!\sum_{k=0}^{3}{{\rm Li}_{4-k}\left(e^{- y_{\rm D}}\right)
\frac{y_{\rm D}^k}{k!}}\nonumber\\&=&
\frac{1}{15}\pi^4+y_{\rm D}^3 \ln \left(1- e^{-y_{\rm D}}\right)\nonumber\\
&-&3y_{\rm D}^2{\rm Li}_{2}\left(e^{-y_{\rm D}}\right) - 6y_{\rm D}{\rm Li}_{3}\left(e^{-y_{\rm D}}\right)-6{\rm Li}_{4}\left(e^{-y_{\rm D}}\right),
\end{eqnarray}
where ${\rm Li}_{s}\left(z\right)=\sum_{k=1}^{\infty}z^k/k^s$ is the polylogarithm of order $s$ and with argument $z$. Substitution of the above formula into Eq.(\ref{eq7}) leads to the final expression:
\begin{eqnarray}
\label{eq11}
F_{\rm D}&=&N\left\{ \frac{9}{8}k_{\rm B}T_{\rm D}-\frac{1}{5}\pi^4 k_{\rm B}T\frac{1}{y_{\rm D}^3} \nonumber   \right. \\ &&\left. + 9k_{\rm B}T\frac{1}{y_{\rm D}}\left[
{\rm Li}_{2}\left(e^{-y_{\rm D}}\right) +\frac{2}{y_{\rm D}}{\rm Li}_{3}\left(e^{-y_{\rm D}}\right)+\frac{2}{y_{\rm D}^2}{\rm Li}_{4}\left(e^{-y_{\rm D}}\right)
\right] \right\}.
\end{eqnarray}

The vibrational energy presented by Eq.(\ref{eq7}) gives also rise to the vibrational pressure:
\begin{equation}
\label{eq12}
p_{\rm D}=-\left(\frac{\partial F_{\rm D}}{\partial V}\right)_T=-\frac{1}{V_0}\left(\frac{\partial F_{\rm D}}{\partial \varepsilon}\right)_T=
9\frac{N}{V_0}k_{\rm B}T_{\rm D}\gamma_{\rm D}       
\left[\frac{1}{8}+\frac{1}{y_{\rm D}^4} \int_{0}^{y_{\rm D}}{\frac{y^3}{e^{y}-1}\,dy}\right] \frac{1}{1+\varepsilon}.
\end{equation}
Again, making use of identity (\ref{eq10}) the Debye contribution to the pressure can be expressed in the form of:
\begin{eqnarray}
\label{eq13}
p_{\rm D}&=&3\frac{N}{V_0}k_{\rm B}T_{\rm D}\gamma_{\rm D}       
\left\{ \frac{3}{8}+\frac{1}{5}\pi^4 \frac{1}{y_{\rm D}^4} +\frac{3}{y_{\rm D}}\ln \left(1- e^{-y_{\rm D}}\right) \right. \nonumber \\ &&\left.
-\frac{9}{y_{\rm D}^2}\left[{\rm Li}_{2}\left(e^{-y_{\rm D}}\right) +\frac{2}{y_{\rm D}}{\rm Li}_{3}\left(e^{-y_{\rm D}}\right)+\frac{2}{y_{\rm D}^2}{\rm Li}_{4}\left(e^{-y_{\rm D}}\right)
\right] \right\} \frac{1}{1+\varepsilon}.
\end{eqnarray}
The formulas (\ref{eq11}) and (\ref{eq13}) describing the vibrational energy and vibrational pressure, respectively, are exact for arbitrary temperature including $T \to 0$ limit. The low-temperature approximation for the Debye model, known from the textbooks, can be obtained with the help of the limiting relationship for polylogarithms: ${\rm lim}_{|z|\to 0}{\rm Li}_s\left(z\right)=z$.

\subsection{The electronic subsystem}

The electronic free energy can be written in the following form \cite{balcerzak_self-consistent_2014}
\begin{equation}
\label{eq14}
F_{\rm el}=N_{\rm el}\left[-\frac{3}{2\pi}\frac{e^2}{4\pi \varepsilon_0}\frac{\sqrt{2m_{\rm el}}}{\hbar}\sqrt{E_{\rm F}}+\frac{3}{5}E_{\rm F}-\frac{\pi^2}{4}\frac{1}{E_{\rm F}}\left(k_{\rm B}T\right)^2\right],
\end{equation}
where $N_{\rm el}$ is the total number of electrons and $E_{\rm F}$ is the Fermi energy. The first term in the formula (\ref{eq14}) corresponds to the exchange energy in the Hartree-Fock approximation \cite{suffczynski_electrons_1985}. The second term is the kinetic energy for $T=0$, and the last term describes thermal energy in the low-temperature region, i.e., when $T\ll E_{\rm F}/k_{\rm B}$. The Fermi energy can be presented as a function of the volume: 
\begin{equation}
\label{eq15}
E_{\rm F}=\frac{\hbar^2}{2m_{\rm el}}\left[3\pi^2\frac{N_{\rm el}}{V} \right]^{2/3}
\end{equation}
which, with the help of Eq.(\ref{eq4}) can be written in the form of:
\begin{equation}
\label{eq16}
E_{\rm F}=E_{0}\left(\frac{r_{0}}{r_{1,0}}\right)^2\frac{1}{\left(1+\varepsilon \right)^{2/3}}
\end{equation}
where
\begin{equation}
\label{eq17}
E_{0}=\frac{\hbar^2}{2m_{\rm el} x^2}\left[3\pi^2z_0\frac{N_{\rm el}}{N} \right]^{2/3}\frac{1}{r_0^2}.
\end{equation}
$E_{0}$ is an energy constant, whereas $(r_{1,0}/r_0)$ ratio (appearing also in the elastic energy) should be found from minimization of the total Gibbs potential at $T=0$, $p=0$ and $H^{z}=0$, i.e., for a non-deformed system. With the help of Eq.(\ref{eq16}) the electronic free energy can be finally expressed in the form:
\begin{equation}
\label{eq18}
F_{\rm el}=N_{\rm el}\left[A E_{0}\left(\frac{r_{0}}{r_{1,0}}\right)\frac{1}{\left(1+\varepsilon \right)^{1/3}}+\frac{3}{5}E_{0}\left(\frac{r_{0}}{r_{1,0}}\right)^2\frac{1}{\left(1+\varepsilon \right)^{2/3}}-\frac{\pi^2}{4}\frac{1}{E_{0}}\left(\frac{r_{1,0}}{r_{0}}\right)^2\left(1+\varepsilon \right)^{2/3}\left(k_{\rm B}T\right)^2\right],
\end{equation}
where the dimensionless $A$-constant is defined as:
\begin{equation}
\label{eq19}
A=-\frac{3}{h}\frac{e^2}{4\pi \varepsilon_0}\sqrt{2m_{\rm el}}\frac{1}{\sqrt{E_{0}}}.
\end{equation}
From the expression (\ref{eq18}) the electronic part of the pressure can be found in the form of:
\begin{eqnarray}
\label{eq20}
p_{\rm el}&=&-\left(\frac{\partial F_{\rm el}}{\partial V}\right)_T = -\frac{1}{V_0}\left(\frac{\partial F_{\rm el}}{\partial \varepsilon}\right)_T= \frac{1}{3}\frac{N_{\rm el}}{V_0}A E_{0}\left(\frac{r_{0}}{r_{1,0}}\right)\frac{1}{\left(1+\varepsilon \right)^{4/3}}
\nonumber \\
&+&\frac{2}{5}\frac{N_{\rm el}}{V_0}E_{0}\left(\frac{r_{0}}{r_{1,0}}\right)^2\frac{1}{\left(1+\varepsilon \right)^{5/3}}+\frac{\pi^2}{6}\frac{N_{\rm el}}{V_0}\frac{1}{E_{0}}\left(\frac{r_{1,0}}{r_{0}}\right)^2\frac{1}{\left(1+\varepsilon \right)^{1/3}}\left(k_{\rm B}T\right)^2,
\end{eqnarray}
which is valid for $T\ll E_{\rm F}/k_{\rm B}$.

\subsection{The magnetic subsystem}

We will consider the magnetic subsystem with the long-range RKKY interaction, as well as with NN direct interaction and the external magnetic field $H^z$ taken into account. We assume that the localized spins have the magnitude $S=1/2$ and they are distributed over a bipartite lattice which consists of two ($a$ and $b$) inter-penetrating sublattices. Having such a lattice we can take into account not only ferromagnetic but also various antiferromagnetic phases \cite{morrish_physical_1965}. We assume that all lattice sites are occupied by the localized spins and no magnetic dilution takes place. In a more general case, where the site dilution and arbitrary spin value $S$ are considered, the magnetic theory in the molecular field approximation (MFA) has been developed in Ref.\cite{szalowski_phase_2008}. That theory can be easily adopted for the present case. In MFA approximation the Gibbs energy of the magnetic subsystem considered here can be presented in the form:
\begin{eqnarray}
\label{eq21}
G_{m}&=&\frac{N}{4}\sum_{k}^{}{\!J_{k}z^{\uparrow\uparrow}_{k}\,\left[\left(m_a\right)^{2}+\left(m_b\right)^{2}\right]}
\!+\frac{N}{2}\sum_{k}^{}{\!J_{k}z^{\uparrow\downarrow}_{k}\,m_a m_b}\nonumber\\
&-&\frac{N}{2}k_{\mathrm{B}}T\mathrm{ln}\left\{2\, \mathrm{cosh}\left[\frac{1}{2k_{\mathrm{B}}T}\left(\Lambda_a+H \right)\right]\right\} \!-\frac{N}{2}k_{\mathrm{B}}T\mathrm{ln}\left\{2\,\mathrm{cosh}\left[\frac{1}{2k_{\mathrm{B}}T}\left(\Lambda_b+H \right)\right]\right\},
\end{eqnarray}
where $J_{k}$ is the exchange integral between the central spin and any spin situated on the $k$-th coordination zone. It is assumed that $J_{k}=J_{k}^{\rm RKKY}$ (for $k=2,3,\ldots$) and $J_{1}=J^{\rm d}+J_{1}^{\rm RKKY}$, whereas $J^{\rm d}$ is the NN direct exchange interaction and $J_{k}^{\rm RKKY}$ (for $k=1,2, \ldots$) is the RKKY interaction. Both these interactions are volume-dependent and their explicit form are given in the Appendix (see eqs.(\ref{A9}) and (\ref{A12})). 

In Eq.(\ref{eq21}) $z^{\uparrow\uparrow}_{k}$ ( $z^{\uparrow\downarrow}_{k}$ ) are the numbers of lattice sites on the $k$-th coordination zone, whose spins are oriented parallel (antiparallel) to the central spin, i.e., they belong to the same (another) sublattice. These numbers satisfy the condition $z^{\uparrow\uparrow}_{k}+z^{\uparrow\downarrow}_{k}=z_{k}$, and their distribution upon $k$ depends on the lattice symmetry and the type of magnetic ordering (ferromagnetic or various antiferromagnetic ones). For each particular case these numbers can be found numerically, by the computer analysis of the given magnetic structure.

In Eq.(\ref{eq21}) $m_i$ (for $i=a,b$) is the $i$-th sublattice magnetization, defined as the thermodynamic mean value of the local spin: $m_i=\langle S_i^z\rangle$. These magnetizations can be found from the set of equations of state (see next subsection).
The exchange fields in Eq.(\ref{eq21}), $\Lambda_i$, acting on the $i$-th sublattice ($i=a,b$) can be expressed by the formulas:
\begin{equation}
\label{eq22}
\Lambda_{a}=\sum_{k}^{}{\!J_{k}\,\left( z^{\uparrow\uparrow}_{k} m_a +z^{\uparrow\downarrow}_{k}m_b\right)}
\;\;\;\;\;\;\; {\rm and} \;\;\;\;\;\;\; 
\Lambda_{b}=\sum_{k}^{}{\!J_{k}\,\left( z^{\uparrow\uparrow}_{k} m_b +z^{\uparrow\downarrow}_{k}m_a\right)}.
\end{equation}
The parameter $H$ is connected with the external magnetic field $H^z$ oriented along $z$-direction and is defined by $H=-g^{\mathrm {eff}}\mu_{\mathrm B}H^{z}$, where $g^{\mathrm {eff}}$ is the effective gyromagnetic factor which for the case of RKKY interaction has been introduced in Ref.~\cite{balcerzak_rkky_2006}. Its explicit form is presented in the Appendix (see Eq.(\ref{A16})). It should be mentioned here that $g^{\mathrm {eff}}$ is volume-dependent (in addition to $J_k$), and this fact has a straightforward consequence for the magnetic pressure calculations.

The magnetic contribution to the pressure can be found from differentiation of Eq.(\ref{eq21}) over the volume:
\begin{eqnarray}
\label{eq23}
p_{m}&=&-\left(\frac{\partial G_{m}}{\partial V}\right)_T = -\frac{1}{V_0}\left(\frac{\partial G_{m}}{\partial \varepsilon}\right)_T = -\frac{1}{2}\frac{N}{V_0}\left(m_a+m_b\right)\mu_{\mathrm B}H^{z}\left(\frac{\partial g^{\mathrm {eff}}}{\partial \varepsilon} \right)
\nonumber \\
&+&\frac{1}{2}\frac{N}{V_0}\sum_{k}^{}\left\{ \frac{1}{2}z^{\uparrow\uparrow}_{k}\,\left[\left(m_a\right)^{2}+\left(m_b\right)^{2}\right]+z^{\uparrow\downarrow}_{k}\,m_a m_b\right\} \left(\frac{\partial J_{k}}{\partial \varepsilon} \right).
\end{eqnarray}
In derivation of Eq.(\ref{eq23}) we already made use of the magnetic equations of state  (see next subsection). The explicit formulas for the derivatives $\partial J_{k}/ \partial \varepsilon$ and $\partial g^{\mathrm {eff}}/ \partial \varepsilon$ are too long to be presented here and are placed in the Appendix (see eqs.(\ref{A11}), (\ref{A13}) and (\ref{A17})).

\subsection{The equations of state}

Having the free energies for all subsystems, given by eqs.(\ref{eq5}), (\ref{eq11}),  (\ref{eq18}) and 
(\ref{eq21}), the total Gibbs energy can be found from Eq.(\ref{eq1}). Then, the set of three equations of state can be obtained by minimizing the total energy with respect to the volume deformation $\varepsilon$ and two magnetizations, $m_a$ and $m_b$, treated as independent variational parameters. Namely, from the condition 
\begin{equation}
\label{eq24}
\frac{\partial G}{\partial \varepsilon}=0,
\end{equation}
we obtain the first equation of state:
\begin{equation}
\label{eq25}
p_{\varepsilon}+p_{\rm D}+p_{\rm el}+p_{m}=p,
\end{equation}
where $p$ is the external pressure, and the subsystem pressures, $p_{\varepsilon}$, $p_{\rm D}$, $p_{\rm el}$ and $p_{m}$, are given by eqs.(\ref{eq6}), (\ref{eq13}), (\ref{eq20}) and (\ref{eq23}), respectively. Two other equations of state are obtained from the requirements that:
\begin{equation}
\label{eq26}
\frac{\partial G}{\partial m_a}=0
\;\;\;\;\;\;\; {\rm and} \;\;\;\;\;\;\; \frac{\partial G}{\partial m_b}=0.
\end{equation}
From these conditions we obtain the following magnetic equations of state:
\begin{equation}
\label{eq27}
m_a=\frac{1}{2} \mathrm{tanh}\left[\frac{1}{2k_{\mathrm{B}}T}\left(\Lambda_a+H \right) \right],
\end{equation} 
and
\begin{equation}
\label{eq28}
m_b=\frac{1}{2} \mathrm{tanh}\left[\frac{1}{2k_{\mathrm{B}}T}\left(\Lambda_b+H \right) \right],
\end{equation} 
where $\Lambda_a$ and $\Lambda_b$ are given by Eq.(\ref{eq22}). The set of three equations of state (\ref{eq25}, \ref{eq27} and \ref{eq28}) should be solved simultaneously for given model parameters, as well as for given temperature $T$ and independent external forces $p$ and $H^z$. Among possible formal solutions obtained for $m_a$, $m_b$ and $\varepsilon$, which may correspond to various magnetically ordered phases, we select that physical solution which gives the minimum of the total Gibbs energy $G$. In particular, for a non-deformed system (NDS), i.e., for $\varepsilon =0$ at $T=0$, $p=0$ and $H^z=0$, the magnetic equations of state are reduced to the limiting values: $m_a=1/2$ and $m_b =\pm 1/2$, where "$+$" sign corresponds to the ferromagnetic phase and "$-$" sign is valid for the antiferromagnetic ones. Then, the first equation of state (\ref{eq22}) can be written as:
\begin{equation}
\label{eq29}
\left( p_{\varepsilon}+p_{\rm D}+p_{\rm el}+p_{m}\right)_{\rm NDS}=0,
\end{equation}
and describes the system in equilibrium for $T=0$, without external forces. From this equation the equilibrium NN distance, $r_{1,0}/r_{0}$, can be found for different magnetic phases. Then, the energy minimum criterion, based on the Gibbs potential, helps to decide which phase is physical in the ground state.

On the other hand, the phase transition temperature $T_{\rm c}$ can be found from linearization of eqs.(\ref{eq27}) and (\ref{eq28}), i.e., when $H^z=0$, $m_a \to 0$ and $m_b \to 0$. This leads to the formula \cite{szalowski_phase_2008}
\begin{equation}
\label{eq30}
k_{\mathrm{B}}T_{\mathrm{c}} = \frac{1}{4}\,\sum_{k}^{}{\!J_{k}\,\left( z^{\uparrow\uparrow}_{k} \pm z^{\uparrow\downarrow}_{k}\right)},
\end{equation}
where the solution with "$+$" corresponds to the Curie temperature and is applicable to the ferromagnetic phase transition, whereas the solution with "$-$" corresponds to the N\'{e}el temperature for the antiferromagnetic phases. Since $J_k$ is volume-dependent, Eq.(\ref{eq30}) allows to study the phase transition (critical) temperature as a function of the external pressure, whereas the volume deformation, $\varepsilon (T_{\rm{c}})$, can be found from the first equation of state (Eq.(\ref{eq25})).

%section 3
\section{Numerical results and discussion}
\label{sec3}

\begin{figure}[h]
  \begin{center}
\includegraphics[width=0.7\columnwidth]{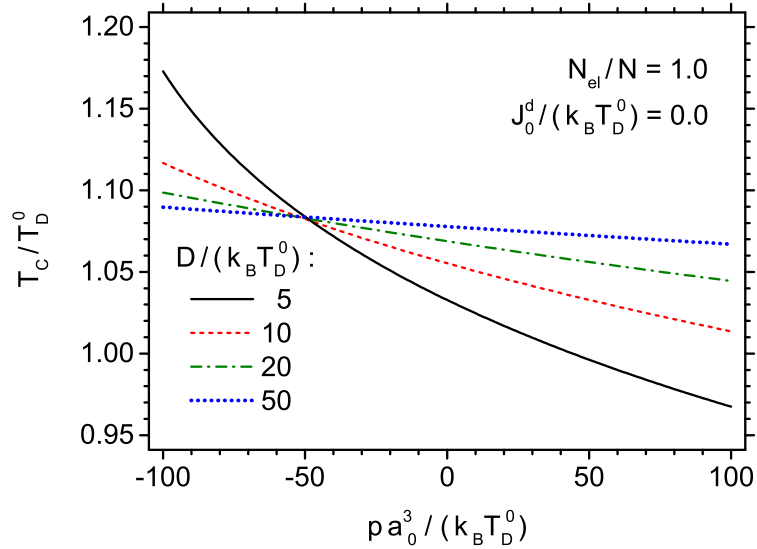}
  \end{center}
   \caption{\label{Fig1}The critical (N\'eel) temperature, $\frac{T_c}{T_{\rm D}^0}$, of phase transition between antiferromagnetic (AF1) and paramagnetic phase, for $\frac{N_{\rm el}}{N}=1$ and $\frac{J_0^d}{k_{\rm B}T_{\rm D}^0}=0$, as a function of the external pressure $\frac{a_0^3}{k_{\rm B}T_{\rm D}^0}p$. Different curves correspond to various depth $\frac{D}{k_{\rm B}T_{\rm D}^0}$ of the Morse potential.}
\end{figure}

\begin{figure}[h]
  \begin{center}
\includegraphics[width=0.7\columnwidth]{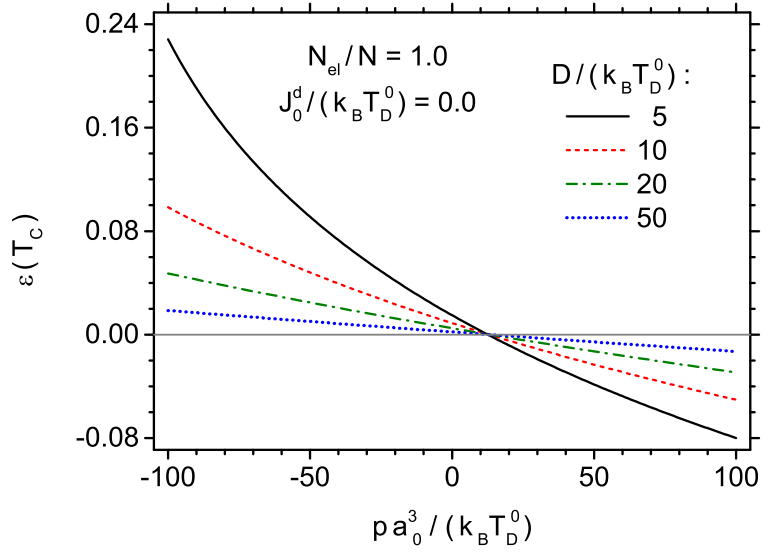}
  \end{center}
   \caption{\label{Fig2}The volume deformation, $\varepsilon (T_c)$, at the critical temperature as a function of the dimensionless external pressure $\frac{a_0^3}{k_{\rm B}T_{\rm D}^0}p$. All remaining parameters are the same as in Fig.~\ref{Fig1}.}
\end{figure}

\begin{figure}[h]
  \begin{center}
\includegraphics[width=0.7\columnwidth]{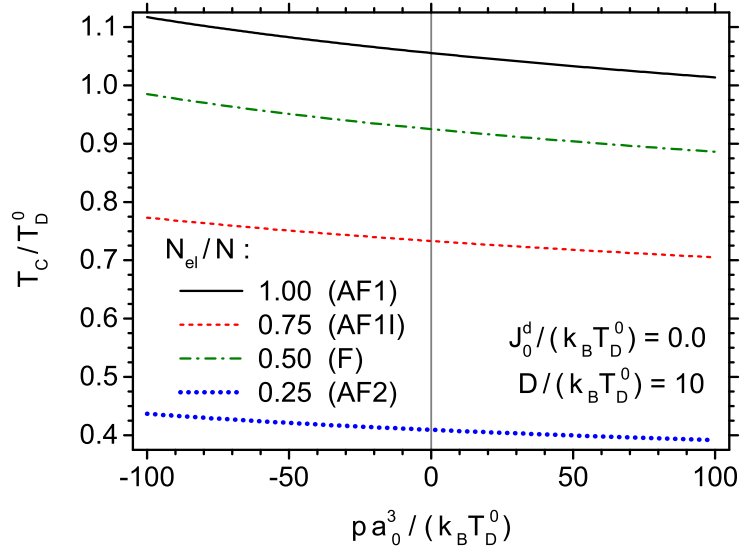}
  \end{center}
   \caption{\label{Fig3}The critical temperature, $\frac{T_c}{T_{\rm D}^0}$, between magnetically ordered and paramagnetic phases, as a function of the external pressure $\frac{a_0^3}{k_{\rm B}T_{\rm D}^0}p$. Different curves correspond to various electron concentrations $\frac{N_{\rm el}}{N}$ and, in consequence, to various magnetic phases. In this figure $\frac{D}{k_{\rm B}T_{\rm D}^0}=10$ and $\frac{J_0^d}{k_{\rm B}T_{\rm D}^0}=0$.}
\end{figure}

\begin{figure}[h]
  \begin{center}
\includegraphics[width=0.7\columnwidth]{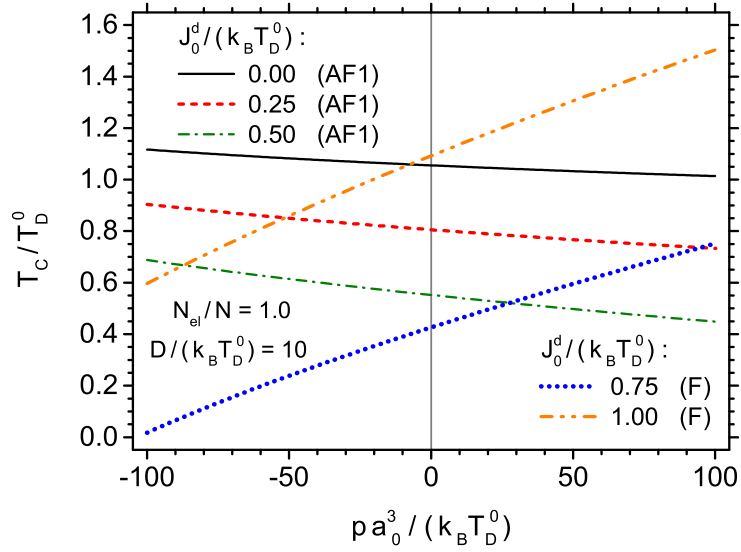}
  \end{center}
   \caption{\label{Fig4}The critical temperature, $\frac{T_c}{T_{\rm D}^0}$, as a function of the external pressure $\frac{a_0^3}{k_{\rm B}T_{\rm D}^0}p$, for various NN exchange interaction parameters $\frac{J_0^d}{k_{\rm B}T_{\rm D}^0}$. The electron concentration is $\frac{N_{\rm el}}{N}=1$, and the Morse potential depth amounts to $\frac{D}{k_{\rm B}T_{\rm D}^0}=10$.}
\end{figure}

\begin{figure}[h]
  \begin{center}
\includegraphics[width=0.7\columnwidth]{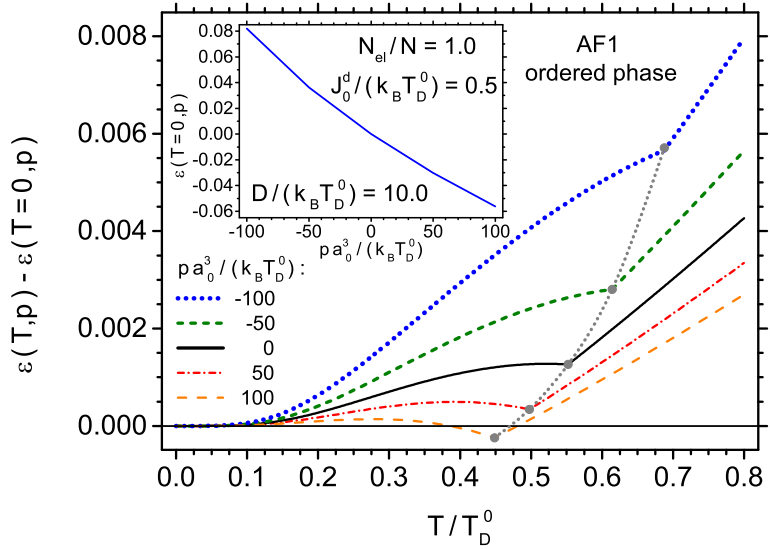}
  \end{center}
   \caption{\label{Fig5}The temperature increase of isotropic volume deformation, $\varepsilon(T,p)-\varepsilon(T=0,p)$, vs. temperature $\frac{T}{T_{\rm D}^0}$, for antiferromagnetic (AF1) phase, when $\frac{N_{\rm el}}{N}=1$. Different curves correspond to various external pressures $\frac{a_0^3}{k_{\rm B}T_{\rm D}^0}p$. The remaining parameters are: $\frac{D}{k_{\rm B}T_{\rm D}^0}=10$ and $\frac{J_0^d}{k_{\rm B}T_{\rm D}^0}=0.5$. By the points the deformations $\varepsilon (T_c/T_{\rm D}^0)$ at the critical temperatures are marked. The inset shows the isotherm for $T = 0$.}
\end{figure}

\begin{figure}[h]
  \begin{center}
\includegraphics[width=0.7\columnwidth]{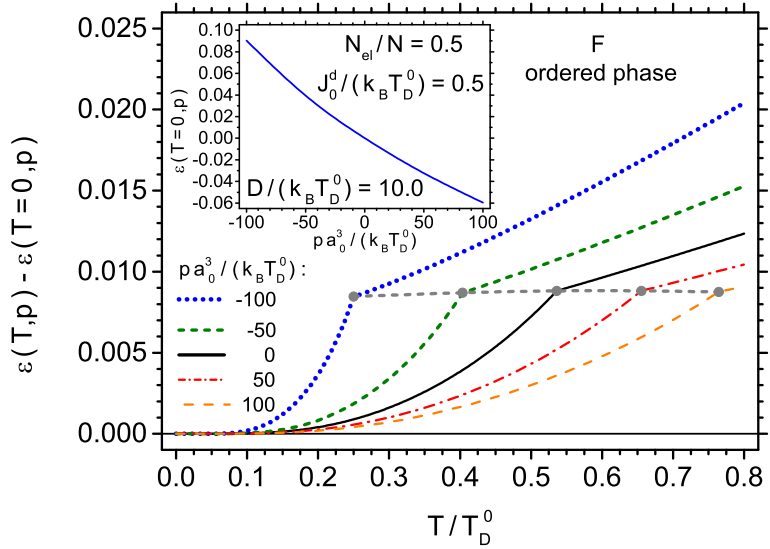}
  \end{center}
   \caption{\label{Fig6}The temperature increase of isotropic volume deformation, $\varepsilon(T,p)-\varepsilon(T=0,p)$, vs. temperature $\frac{T}{T_{\rm D}^0}$, for ferromagnetic  phase, when $\frac{N_{\rm el}}{N}=0.5$. Different curves correspond to various external pressures $\frac{a_0^3}{k_{\rm B}T_{\rm D}^0}p$. The remaining parameters are the same as in Fig.~\ref{Fig5}. By the points the deformations $\varepsilon (T_c/T_{\rm D}^0)$ at the critical temperatures are marked. The inset shows the isotherm for $T = 0$.}
\end{figure}

\begin{figure}[h]
  \begin{center}
\includegraphics[width=0.7\columnwidth]{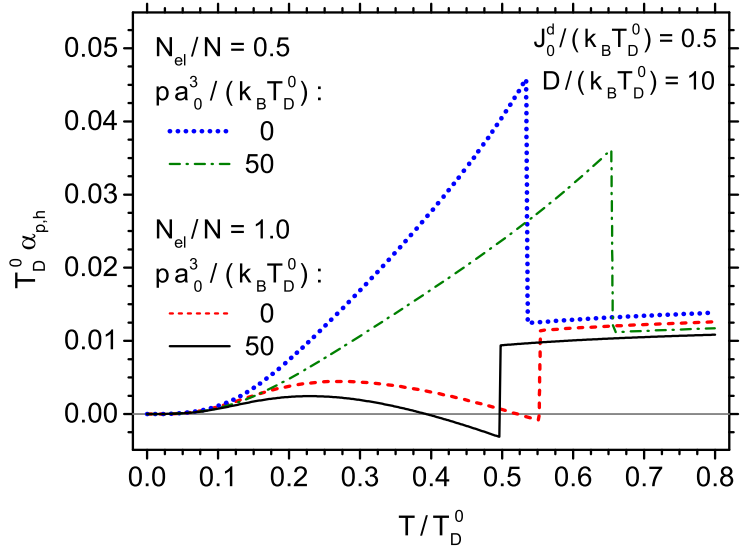}
  \end{center}
   \caption{\label{Fig7}The thermal volume expansion coefficient,  $\alpha_{p,h}$, in dimensionless units, vs. temperature $\frac{T}{T_{\rm D}^0}$, for $\frac{D}{k_{\rm B}T_{\rm D}^0}=10$ and $\frac{J_0^d}{k_{\rm B}T_{\rm D}^0}=0.5$. Different curves correspond to two electron concentrations: $\frac{N_{\rm el}}{N}=1$ and $\frac{N_{\rm el}}{N}=0.5$, as well as to two different pressures: $p=0$ and $\frac{a_0^3}{k_{\rm B}T_{\rm D}^0}p=50$.}
\end{figure}

\begin{figure}[h]
  \begin{center}
\includegraphics[width=0.7\columnwidth]{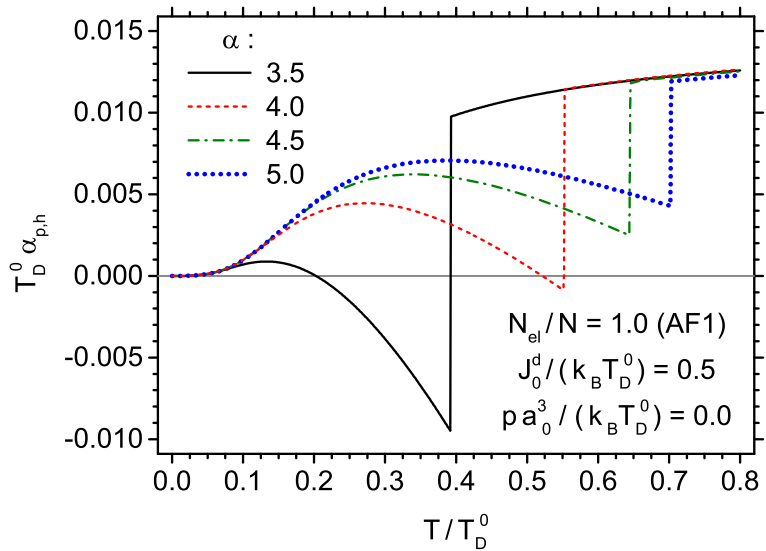}
  \end{center}
   \caption{\label{Fig8}The thermal volume expansion coefficient,  $\alpha_{p,h}$, in dimensionless units, vs. temperature $\frac{T}{T_{\rm D}^0}$, for antiferromagnetic (AF1) phase, when $\frac{D}{k_{\rm B}T_{\rm D}^0}=10$, $\frac{J_0^d}{k_{\rm B}T_{\rm D}^0}=0.5$, $\frac{N_{\rm el}}{N}=1$ and $p=0$. Different curves illustrate the influence of the asymmetry parameter $\alpha$ of the Morse potential.}
\end{figure}

\begin{figure}[h]
  \begin{center}
\includegraphics[width=0.7\columnwidth]{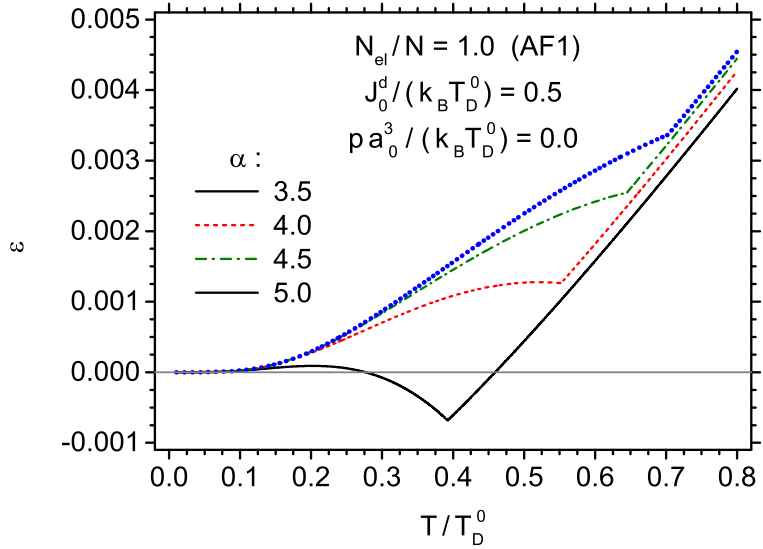}
  \end{center}
   \caption{\label{Fig9}The isotropic volume deformation, $\varepsilon$, vs. temperature $\frac{T}{T_{\rm D}^0}$, for antiferromagnetic (AF1) phase, when the rest of parameters are the same as in Fig.~\ref{Fig8}. Different curves illustrate the influence of the asymmetry parameter $\alpha$ of the Morse potential.}
\end{figure}

\begin{figure}[h]
  \begin{center}
\includegraphics[width=0.7\columnwidth]{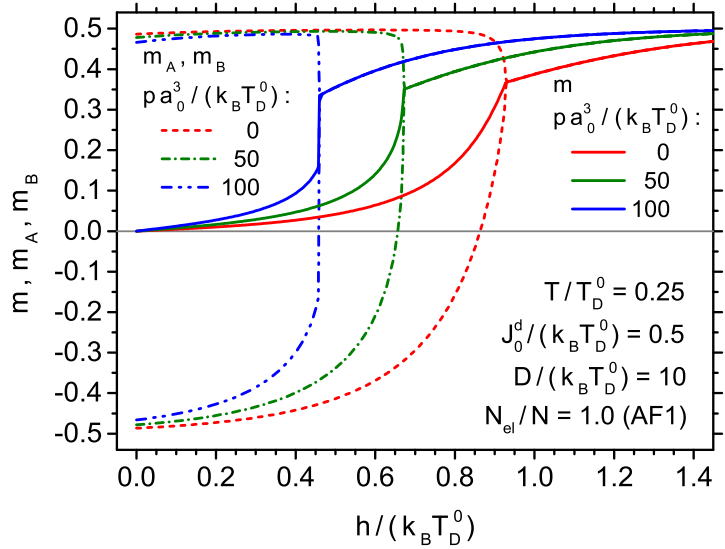}
  \end{center}
   \caption{\label{Fig10}The sublattice magnetizations, $m_a$ and $m_b$, as well as the total magnetization $m$ of antiferromagnetic (AF1) phase vs. dimensionless external magnetic field, $-\frac{g_S\mu_{\mathrm B}H^{z}}{k_{\rm B}T_{\rm D}^0}$, for $\frac{T}{T_{\rm D}^0}=0.25$. Different curves correspond to various external pressures $\frac{a_0^3}{k_{\rm B}T_{\rm D}^0}p$. The remaining parameters are: $\frac{D}{k_{\rm B}T_{\rm D}^0}=10$, $\frac{J_0^d}{k_{\rm B}T_{\rm D}^0}=0.5$ and $\frac{N_{\rm el}}{N}=1$.}
\end{figure}

\begin{figure}[h]
  \begin{center}
\includegraphics[width=0.7\columnwidth]{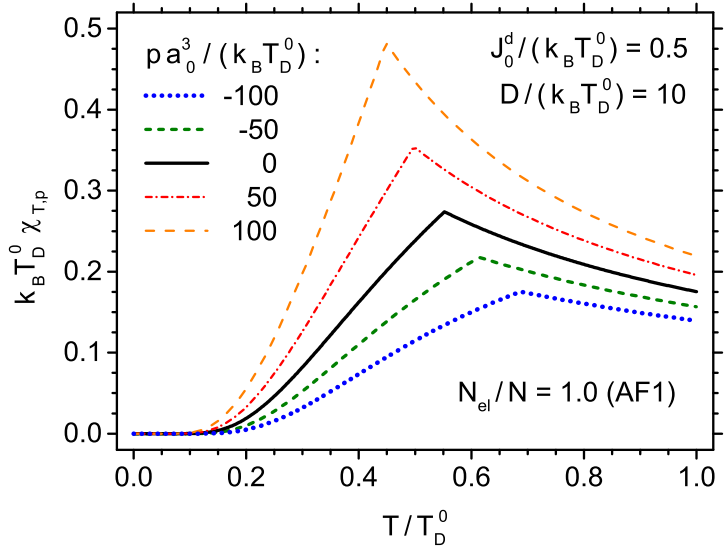}
  \end{center}
   \caption{\label{Fig11}The magnetic susceptibility, $\chi_{T,p}$, in dimensionless units, vs. temperature $\frac{T}{T_{\rm D}^0}$, for antiferromagnetic (AF1) phase, when $\frac{N_{\rm el}}{N}=1$. Different curves correspond to various external pressures $\frac{a_0^3}{k_{\rm B}T_{\rm D}^0}p$. The remaining parameters are: $\frac{D}{k_{\rm B}T_{\rm D}^0}=10$ and $\frac{J_0^d}{k_{\rm B}T_{\rm D}^0}=0.5$.}
\end{figure}

\begin{figure}[h]
  \begin{center}
\includegraphics[width=0.7\columnwidth]{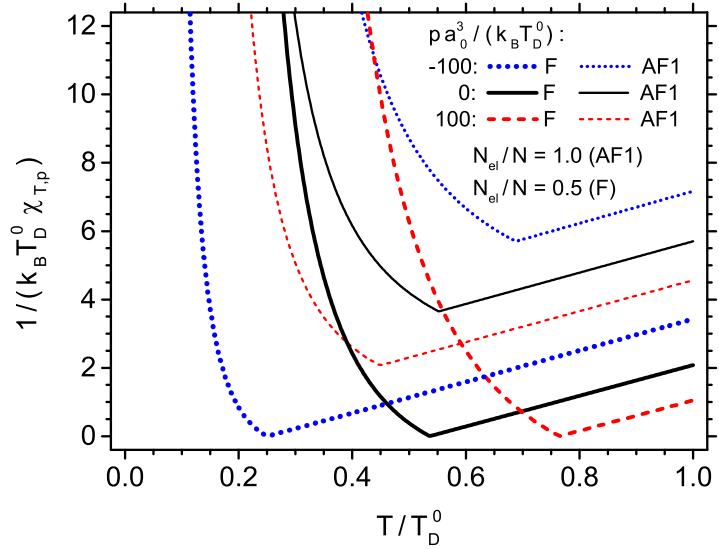}
  \end{center}
   \caption{\label{Fig12}The inverse of magnetic susceptibility, $\chi_{T,p}$, in dimensionless units, vs. temperature $\frac{T}{T_{\rm D}^0}$, for ferromagnetic  phase, when $\frac{N_{\rm el}}{N}=0.5$, and for antiferromagnetic phase, when $\frac{N_{\rm el}}{N}=1.0$. Different curves correspond to various external pressures $\frac{a_0^3}{k_{\rm B}T_{\rm D}^0}p$. The remaining parameters are the same as in Fig.~\ref{Fig11}.}
\end{figure}

\begin{figure}[h]
  \begin{center}
\includegraphics[width=0.7\columnwidth]{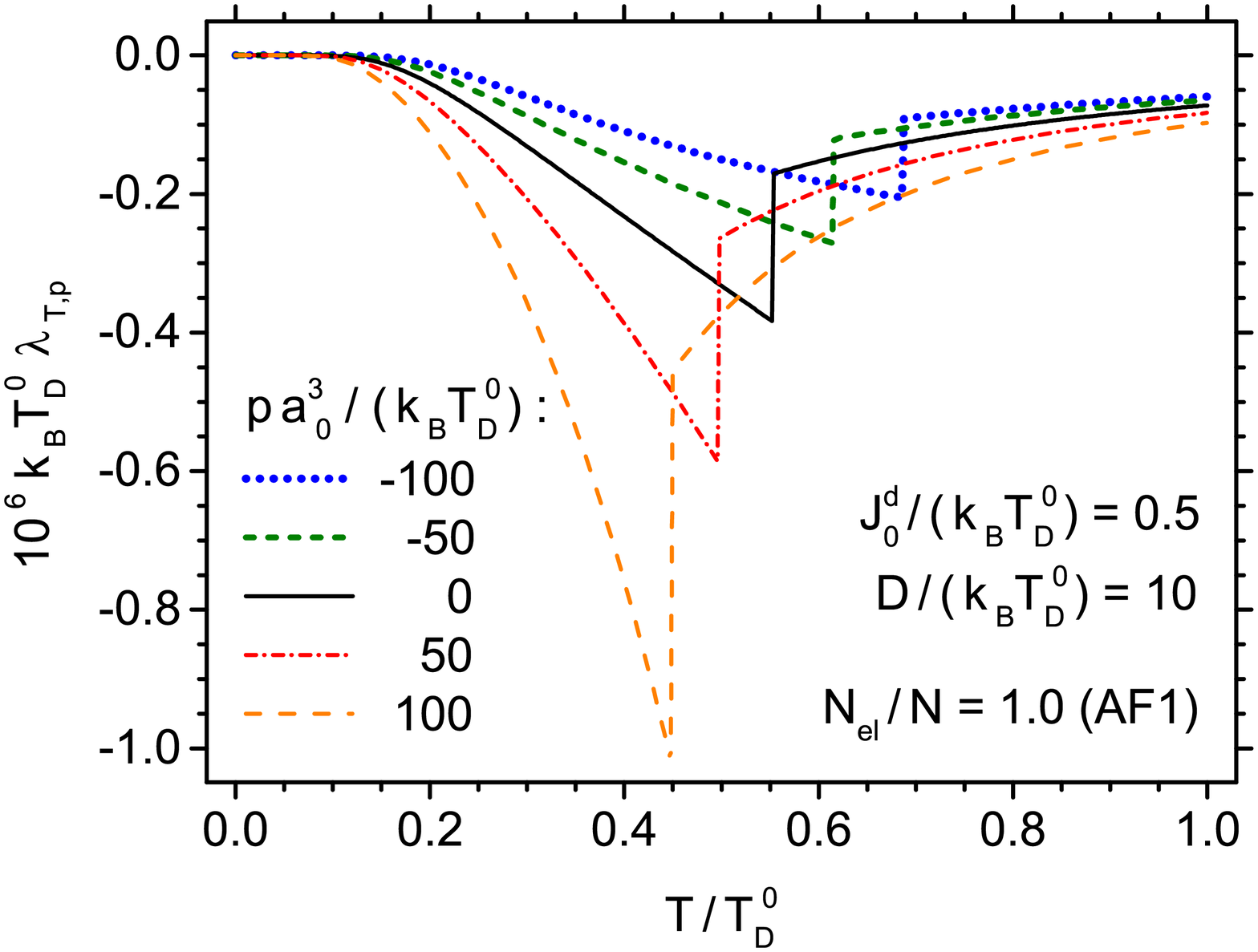}
  \end{center}
   \caption{\label{Fig13}The magnetostriction coefficient, $\lambda_{T,p}$, in dimensionless units, as a function of temperature $\frac{T}{T_{\rm D}^0}$, for antiferromagnetic (AF1) phase, when $\frac{N_{\rm el}}{N}=1$. Different curves correspond to various external pressures $\frac{a_0^3}{k_{\rm B}T_{\rm D}^0}p$. The remaining parameters are: $\frac{D}{k_{\rm B}T_{\rm D}^0}=10$ and $\frac{J_0^d}{k_{\rm B}T_{\rm D}^0}=0.5$.}
\end{figure}

The numerical results will be presented for a model lattice with FCC structure and some exemplary, typical interaction parameters. For FCC lattice we have $x=\sqrt{2}$ and $z_0=4$. The energy constants will be expressed in $k_{\rm B}T_{\rm D}^0$ units, where $T_{\rm D}^0$ is the Debye temperature of a non-deformed system. For the Morse potential we assume the asymmetry parameter mostly as $\alpha=4$, whereas the normalized potential depth, $\frac{D}{k_{\rm B}T_{\rm D}^0}$, is accepted with the values of 10.  The $q$-parameter, appearing in Eq.(\ref{eq8}) for the Debye temperature, is equal to $q=1$. Regarding the RKKY interaction (see Appendix), we assume that $\frac{C_0}{k_{\rm B}T_{\rm D}^0}=0.5$, $\frac{C_0}{J}= -0.005$ and $\frac{g_{\rm el}}{g_S}=\frac{4}{3}$. The damping effect can be neglected here, $\lambda \to \infty$, since we deal with a pure crystal. For the NN interactions, the exponent $n$, appearing in Eq.(\ref{A12}), is assumed with the value $n=6$, analogously to Ref.\cite{balcerzak_self-consistent_2017}. As far as the electron subsystem is concerned, the energy parameter $E_0$ from Eq.(\ref{eq17}) amounts to $E_0=E_0^{\prime}\left(\frac{N_{\rm el}}{N} \right)^{2/3}$, and we adopt the value 
$\frac{E_0^{\prime}}{k_{\rm B}T_{\rm D}^0}=200$ for the numerical calculations. By the same token, $A=\frac{A^{\prime}}{\left(N_{\rm el}/N \right)^{1/3}}$ in Eq.(\ref{eq19}), and the assumed numerical value of $A^{\prime}$ is: $A^{\prime}=-1.6$. In turn, the electron density, $\frac{N_{\rm el}}{N}$, can be freely varied. It determines both the electron gas properties and the magnetic RKKY interaction characteristic as well. In the presentation of numerical results we will use the dimensionless absolute temperature, $\frac{T}{T_{\rm D}^0}$, the dimensionless external pressure, $\frac{a_0^3}{k_{\rm B}T_{\rm D}^0}p$ (where $a_0$ is the lattice constant of NDS), and the dimensionless external magnetic field $\frac{h}{k_{\rm B}T_{\rm D}^0}$, where $h=-g_S\mu_{\mathrm B}H^{z}$, oriented along $z$ direction.\\

The RKKY interaction, supplemented with the NN direct exchange, can lead to the spontaneous magnetic ordering which, depending on the electron concentration, is either ferromagnetic or antiferromagnetic. In case of FCC structure, three antiferromagnetic phases are possible, and they have been classified in Ref.\cite{morrish_physical_1965} as: antiferromagnetic first kind (AF1), antiferromagnetic first kind improved (AF1I) and antiferromagnetic second kind (AF2). According to the atomic arrangements on $a$ and $b$ sublattices, as presented in Ref.\cite{morrish_physical_1965}, for each magnetic phase the coordination numbers $z^{\uparrow\uparrow}_{k}$ and  $z^{\uparrow\downarrow}_{k}$ ($k$=1, 2, ..., $k_{\rm max}$) have been found numerically. For the long-range RKKY interaction we found that a satisfactory convergence of the magnetic energy was achieved when the maximum number of coordination zones is $k_{\rm max}=41253$, which corresponds to 150 lattice constants. On the other hand, for the Morse potential a satisfactory convergence was obtained when $k_{\rm max}=265$, which corresponds to 12 lattice constants. Having found all the coordination numbers, $z^{\uparrow\uparrow}_{k}$ and  $z^{\uparrow\downarrow}_{k}$, as well as the exchange interactions calculated for each coordination zone $k$, we solve the equation (\ref{eq29}) for NN equilibrium distances for NDS (i.e., $r_{1,0}/r_{0}$) for each magnetic phase and we calculate the corresponding total Gibbs energies. From the minimum criterion for the total energy, the physical ground state is obtained. It contains information about the equilibrium distance, $r_{1,0}/r_{0}$, for $T=0$, $p=0$, and $H^z=0$ (whereas $\varepsilon =0$) and specifies the magnetic phase with the lowest energy. On this basis, further numerical calculations can be carried out by solving the set of three equations of state (\ref{eq25}, \ref{eq27} and \ref{eq28}) for arbitrary temperature, external pressure and magnetic field. The set of coupled equations of state yield solutions for the equilibrium deformation $\varepsilon$, as well as the magnetizations $m_a$ and $m_b$, under the influence of external conditions.

At first, we performed numerical calculations when the NN direct exchange was set to zero in order to see the pure RKKY coupling effect on the calculated properties. For some choice of the electron density $N_{\rm el}/N$ the results are presented in Figs.~\ref{Fig1}-\ref{Fig3}. In Fig.~\ref{Fig1}, for $N_{\rm el}/N=1$, the critical (N\'eel) temperature of phase transition between AF1 and paramagnetic phase is presented as a function of the external pressure, for various depth of the Morse potential $D/(k_{\rm B}T_{\rm D}^0)$. For the values $D/(k_{\rm B}T_{\rm D}^0)=$5, 10, 20 and 50, the NN equilibrium distances are: $r_{1,0}/r_{0}=$ 0.875378, 0.886628, 0.893457 and 0.898071, respectively. These values, being noticeably lower than 1, reflect a strong compression of the system, which is due to the exchange interaction in the electron gas described by the Hartree-Fock term. The compressive forces have to be balanced by other pressure contributions, mainly by the strong repulsive forces originating from the Morse potential. The negative external pressure in Fig.~\ref{Fig1} means the presence of stretching forces, whereas the positive pressure corresponds to isotropic, hydrostatic compressive pressure. Let us mention that the stretching forces can also originate from the internal, molecular pressure, as a result of crystal doping, with the atoms which are located in interstitial positions. In general, the critical temperature of antiferromagnetic phase decreases when the pressure increases, and the dependency is stronger if the potential becomes more shallow. At some characteristic negative pressure the critical temperature becomes only weakly sensitive to the potential depth.

In Fig.~\ref{Fig2} the volume deformation $\varepsilon (T_c)$ at the critical temperature is shown, for the same parameters as in Fig.~\ref{Fig1}. The negative volume deformation corresponds to the compression of the system. Pressure dependency of the curves in Fig.~\ref{Fig2} is similar to the corresponding curves from Fig.~\ref{Fig1}. In particular, it can be noticed that in the vicinity of some small positive pressure the deformation at critical temperature is reduced to zero and this characteristic pressure is insensitive to the Morse potential depth $D$.

In Fig.\ref{Fig3} the potential depth is fixed at $D/(k_{\rm B}T_{\rm D}^0)=10$, whereas the electron density varies, taking the values $N_{\rm el}/N=$ 1.5, 1, 0.5 and 0.25, which correspond to the existence of ferromagnetic, AF1, AF1I and AF2 phases, respectively. Then, the appropriate NN equilibrium distances are: $r_{1,0}/r_{0}=$ 0.895425, 0.886628, 0.888981 and 0.894928, respectively. The critical temperatures, shown in Fig.~\ref{Fig3}, diminish with an increase of pressure, and their values strongly depend on the kind of magnetic phase. We note that the case of $N_{\rm el}/N=1$ corresponds to AF1 phase from Figs.~\ref{Fig1} and \ref{Fig2}.

We found that by introducing the NN direct exchange integral, in addition to the long-range RKKY interaction, the magnetic properties can be markedly modified. In Fig.~\ref{Fig4} the influence of NN interaction parameter, $J_0^d/(k_{\rm B}T_{\rm D}^0)$, on the phase transition temperature dependencies vs. external pressure is presented. The electron density is fixed in this case at $N_{\rm el}/N=1$ and the potential depth is $D/(k_{\rm B}T_{\rm D}^0)=10$. It is seen that depending on the $J_0^d/(k_{\rm B}T_{\rm D}^0)$ value the kind of magnetic ordering can be changed from AF1 phase to the ferromagnetic one, whereas $J_0^d/(k_{\rm B}T_{\rm D}^0)$ changes from 0 to 1. Moreover, in the ferromagnetic phase the critical (Curie) temperature is now an increasing function vs. external pressure, in contrast to previous results for $J_0^d=0$. Such an increasing character of $T_c$ vs. $p$ has already been found in our previous paper \cite{balcerzak_self-consistent_2017}, where only NN ferromagnetic interaction was present. 

In further presentation of the numerical results we will restrict ourselves to the NN interaction value $J_0^d/(k_{\rm B}T_{\rm D}^0)=0.5$ and $D/(k_{\rm B}T_{\rm D}^0)=10$, whereas the electron density will be selected as either $N_{\rm el}/N=1$ or $N_{\rm el}/N=0.5$. For $N_{\rm el}/N=1$ the antiferromagnetic (AF1) phase is present, and the NN equilibrium distance is $r_{1,0}/r_{0}=$ 0.887019. On the other hand, for $N_{\rm el}/N=0.5$, we deal with the ferromagnetic phase where $r_{1,0}/r_{0}=$ 0.887201.

The temperature dependencies of the volume deformation, $\varepsilon$, for $N_{\rm el}/N=1$ (AF1 phase) and for $N_{\rm el}/N=0.5$ (ferromagnetic F phase) are illustrated in Fig.~\ref{Fig5} and Fig.~\ref{Fig6}, respectively. In these plots a few values of normalized external pressure are accepted. In both figures, the main plot shows the volume deformation $\varepsilon$ with its value at $T=0$ subtracted, i.e. $\varepsilon(T,p)-\varepsilon(T=0,p)$ is plotted. This is done in order to separate the sole influence of the temperature from the (larger) effect of external pressure. The sensitivity of $\varepsilon$ to $p$ can be assessed on the basis of insets, which present isotherms corresponding to the dependence of $\varepsilon$ on $p$ at $T=0$. It can be seen that the external pressure (in the studied range) exerts an order of magnitude larger effect on the relative deformation that the temperature. Therefore, plotting $\varepsilon(T,p)-\varepsilon(T=0,p)$ was necessary to emphasize the effect of $T$ at various pressures. Both for $N_e/N=0.5$ and $N_e/N=1.0$, the relative deformation at $T=0$ decreases with the pressure and both isotherms are rather similar. 

In both figures the volume deformations at the phase transition temperatures, i.e., for $T=T_c$, are marked by bold points and connected with the dashed line. 

The fact that the temperature dependence of the volume deformation is relatively weak compared to pressure dependence, can be explained by the strong compressive forces originating from the electron gas subsystem, as discussed previously. The resulting electron pressure pushes the NN equilibrium distance well below $r_0$, where $r_0$ corresponds to a minimum of the Morse potential. It means that the NN distance is moved to the region where this potential strongly depends on $r$. In that region the overall balance of the external pressure with the internal pressure contributions, imposed by the equation of state (\ref{eq25}), can be achieved by only the small changes of interatomic distance.

For both considered concentrations of electrons, above the critical temperature the behaviour of $\varepsilon$ as a function of $T$ is linear, while for $T<T_{c}$ a significant nonlinearity is observed. For AF1 ordered phase and zero pressure, the relative volume deformation increases with the temperature faster than linearly, then the dependence flattens in the vicinity of $T_c$. If the stretching pressure is applied, the increase of $\varepsilon$ with $T$ is faster at low temperatures and the flattening is less pronounced. On the contrary, the compressive pressure reduces the thermal expansion, and even in some range of $T<T_c$ the relative deformation decreases. At $T=T_c$ a kink is observed in the dependencies and it is seen how the positive external pressure reduces the critical temperature, while the stretching pressure acts in the opposite way (see Fig.~\ref{Fig1}).

When the F ordered phase is considered, the volume deformation is always an increasing function of the temperature and it varies faster than linearly for $T<T_c$. The external pressure only shifts the critical temperature (which rises under the action of the compressive pressure, contrary to the AF1 phase case). It is also evident that the change in $\varepsilon$ between $T=0$ and $T=T_c$ is very weakly dependent on the pressure, which also contrasts with the behaviour seen for AF1 ordered phase.

In order to examine the thermal behaviour of the volume, the thermal expansion coefficient can be analysed. We define the volume thermal expansion coefficient, $\alpha_{p,h}$, at constant pressure $p$ and constant magnetic field $h=-g_S\mu_{\mathrm B}H^{z}$, as $\alpha_{p,h}=1/V\left(\partial V/\partial T\right)_{p,h}$, where $V=V_0\left(1+\varepsilon\right)$. In order to obtain $\alpha_{p,h}$, the thermal derivatives of the volume deformation curves $\varepsilon$ from Figs.~\ref{Fig5} and ~\ref{Fig6} are calculated. The thermal expansion coefficient in dimensionless units is presented in Fig.~\ref{Fig7}, simultaneously for AF1 and ferromagnetic (F) phases, and for two selected pressures: $p=0$ and $\frac{a_0^3}{k_{\rm B}T_{\rm D}^0}p=50$. The jumps of $\alpha_{p,h}$ at the critical temperatures are clearly seen for both phases, which is a consequence of magnetoelastic coupling \cite{callen_magnetostriction_1965}. It can be noted that behaviour of $\alpha_{p,h}$ for the AF1 and F phases, is very different. In particular, the directions of the jumps at $T_c$ are opposite for the N\'eel and Curie temperatures. Moreover, it is seen that in some region just below the N\'eel temperature the thermal expansion coefficient becomes negative. 

The negative $\alpha_{p,h}$-coefficient may be considered as a surprising result, but can be explained after analysis of the individual pressure contributions to the equation of state (\ref{eq25}). Namely, it can be shown that in this range of parameters the magnetic contribution to the pressure is negative (i.e., compressive) for F phase, but is positive (i.e., expansive) for AF1 phase. Moreover, the magnetic pressure vanishes at $T=T_c$, and is zero for the paramagnetic phase (see Eq.(\ref{eq23}). It means that when the system approaches the N\'eel temperature from AF1 phase, the magnetic pressure quickly decreases with temperature. For constant external pressure, this decrease must be compensated by an increase of other pressure contributions, in particular by the positive change of the elastic pressure. In fact, an increase of the elastic positive pressure, resulting from the Morse potential, is obtained when the NN distances become shorter, which means the volume compression. On the other hand, for the ferromagnetic phase an analogous effect of compensation is acting in the opposite direction, leading to an increase of $\alpha_{p,h}$-coefficient when the Curie temperature is approached from ferromagnetic phase.

The effect of negative thermal expansion depends on the parameters of the Morse potential. We found that the asymmetry parameter $\alpha$ appearing in Eq.(\ref{eq2}) plays an important role. This is illustrated in Fig.~\ref{Fig8} for AF1 phase, when the external pressure is $p=0$. Different curves correspond to various $\alpha$- asymmetry parameters. The shift of the N\'eel temperature (where the jump occurs) towards lower values is seen when $\alpha$ decreases. The curve labelled by $\alpha =4$ is the same as the corresponding curve from Fig.~\ref{Fig5}. For $\alpha =4$ and 3.5 the negative thermal expansion coefficient is obtained, and the effect is even much stronger for $\alpha =3.5$, i.e., when the Morse potential is more symmetric. This fact is in agreement with our explanation of the effect given above. First of all, when the elastic potential is more symmetric the thermal expansion diminishes. At the same time, the potential is less steep for $r<r_0$, which means that the necessary increase of the elastic pressure, compensating a simultaneous decrease of the magnetic (positive) pressure, can be achieved by some larger shift of NN distance towards lower values. This means a further decrease of the resultant thermal expansion coefficient, which may even enter the negative values if $\alpha$ is small enough.

In order to complement the above discussion, in Fig.~\ref{Fig9} the volume deformation $\varepsilon$ is presented vs. temperature for all parameters being the same as in Fig.~\ref{Fig8}. It is seen that the curves markedly differ for various $\alpha$, and the kinks which occur at the N\'eel temperature are signalling the jumps of the thermal expansion coefficient. For the asymmetry parameter $\alpha=3.5$, the volume deformation is very small in antiferromagnetic phase and almost constant in a wide range of low temperatures. This can be interpreted as the invar-like effect. Interestingly, when $T$ approaches the N\'eel temperature, the volume deformation changes more rapidly and it may even become negative.

In order to study the influence of external magnetic field on the thermodynamic properties, we started with its effect on the magnetization. In Fig.~\ref{Fig10} the sublattice magnetizations of antiferromagnetic phase AF1 is plotted vs. external field at constant temperature $T/T_{\rm D}^0=0.25$, together with the total magnetization, $m=\left(m_A+m_B\right)/2$. There values of the external pressures are chosen: $\frac{a_0^3}{k_{\rm B}T_{\rm D}^0}p=0$, 50 and 100. We see that with an increase of the magnetic field, the spin-flip re-orientation of one sublattice takes place, and eventually both sublattices become parallel ordered, with the same magnitude. The existence of the critical field, at which the magnitudes of both sublattices become the same and the phase transition to the ferromagnetically ordered state takes place, is evident. This critical field depends on the pressure, and decreases when the pressure increases. Interestingly, for $\frac{a_0^3}{k_{\rm B}T_{\rm D}^0}p=100$ such phase transition is found to be of the 1st order. After spin-flip to the ferromagnetic phase, when the external field further increases, magnetization of both sublattices tend towards the saturation value 1/2.

In Fig.~\ref{Fig11}, one of the magnetic response functions, i.e., the susceptibility $\chi_{T,p}$, defined as $\chi_{T,p}=\left(\frac{\partial m}{\partial h}\right)_{T,p}$ is presented for AF1 phase, when $N_{\rm el}/N=1$. In the case of antiferromagnetic ordering, by $m$ we mean the average magnetization: $m=(m_a+m_b)/2$, whereas the external magnetic field is parametrized by $h=-g_S\mu_{\mathrm B}H^{z}$. Different curves in Fig.~\ref{Fig11} correspond to various external pressures. The susceptibility behaves typically for antiferromagnets, with a finite maximum at the N\'eel temperature. As discussed before, the position of the phase transition temperature decreases with an increase of the pressure. Moreover, it is seen that the peak of $\chi_{T,p}$ becomes sharper when $p$ increases.

In turn, the inverse of the magnetic susceptibility is presented in Fig.~\ref{Fig12}, for $N_{\rm el}/N=0.5$ and $N_{\rm el}/N=1.0$, and various pressures. The curves present the ferromagnetic susceptibility (for $N_{\rm el}/N=0.5$) together with the data for the previous case of AF1 phase. In the ferromagnetic case, the susceptibility has a pole at the Curie temperature (the inverse takes the value of zero), as expected. It is also seen that increasing pressure shifts the phase transition point towards higher temperatures, in contrast to the antiferromagnetic phase (compare with Fig.~\ref{Fig11}).

One of the important magnetoelastic phenomena is magnetostriction. The volume magnetostriction coefficient $\lambda_{T,p}$ is defined as $\lambda_{T,p}=\frac{1}{V}\left(\frac{\partial V}{\partial h}\right)_{T,p}$. It is presented as a function of temperature in Fig.~\ref{Fig13} for AF1 phase, when $N_{\rm el}/N=1$, and in Fig.~\ref{Fig14} for ferromagnetic phase, when $N_{\rm el}/N=0.5$. Several curves correspond to various constant pressures. It is seen that the magnetostriction coefficient is negative for both AF1 and F phases and reveals the sharp minima at the phase transition temperatures. However, the shapes of these minima in both phases are different. For instance, in AF1 phase the negative peaks are relatively broad, finite, and their magnitudes increase with an increase of the pressure. On the other hand, in F phase the peaks are very narrow, going to infinity, and, as the pressure increases, they become numerically less pronounced. 

The piezomagnetic effect, defined by the derivative $\left(\frac{\partial m}{\partial p}\right)_{T,h}$ is presented in Fig.~\ref{Fig15} for the ferromagnetic phase.  Different curves correspond to some selected values of the external magnetic field. For $h$=0, when the phase transition takes place, the piezomagnetic coefficient becomes divergent at the Curie temperature. In the presence of the magnetic field the maximum of that coefficient is finite and it tends to be reduced when $h$ increases. At the same time, the maximum broadens and shifts to higher temperatures. It is also seen that in the low temperature region the external pressure $p$ has practically no effect on the magnetization $m$. It should also be mentioned that the piezomagnetic coefficient $\left(\frac{\partial m}{\partial p}\right)_{T,h}$ can be related to the magnetostriction coefficient $\lambda_{T,p}$. Namely, using the general relationship $\frac{\partial^2 G}{\partial h \partial p}=\frac{\partial^2 G}{\partial p \partial h}$ one obtains: $\left(\frac{\partial m}{\partial p}\right)_{T,h}=-\frac{a_0^3}{z_0}\left(1+\varepsilon \right)\lambda_{T,p}$. In this way, one can relate the curve for $H^z=0$ to the curve labelled with "0" in Fig.~\ref{Fig14}.

In the last figure (Fig.~\ref{Fig16}), the isothermal compressibility, $\kappa_{T,h}$, is plotted vs. temperature, both for AF1 phase ($N_{\rm el}/N=1$) and for F phase ($N_{\rm el}/N=0.5$). The isothermal compressibility is defined as $\kappa_{T,h}=-\frac{1}{V}\left(\frac{\partial V}{\partial p}\right)_{T,h}$, and is calculated for $h=0$. Different curves in Fig.~\ref{Fig16} correspond to three values of the external pressure. At the critical temperature a rapid decrease of the isothermal compressibility is observed when the system goes from magnetically ordered to paramagnetic phase. A size of this jump is greater for the ferromagnetic phase than for AF1, and weakly depends on the pressure. However, the pressure influences the phase transition temperature where the jump takes place, and shifts the position of the curves on the vertical scale.

\section{Final remarks}
\label{sec4}

In the paper a self-consistent thermodynamic model of the solid crystal with RKKY interaction and magnetoelastic coupling is presented. The generalized Gibbs potential and the set of equations of state is derived, from which all thermodynamic properties are obtained. For a two sublattice system the equations of state contain 6 variables: isotropic volume deformation $\varepsilon$, sublattice magnetizations $m_a$ and $m_b$, temperature $T$, external pressure $p$ and magnetic field $H^z$. These equations have been solved numerically for a model crystal with FCC structure, treating $T$, $p$ and $H^z$ as independent variables. Because the model possesses very rich possibilities concerning the choice of its parameters, and a variety of properties can be calculated, in presentation of the numerical results only the most characteristic phenomena have been illustrated.

In the paper we studied only solutions for several most common magnetic phases: ferromagnetic, three antiferromagnetic (AF1, AF1I and AF2) as well as the paramagnetic phase. For each phase the Gibbs energy has been constructed and the numerical unique solution has been found using the lowest total energy criterion. Other, more complicated magnetic phases like, for instance, spin-glass phase, have not been studied since it would exceed the frame of the paper.

Since the total energy has been constructed as a sum of the subsystem energies, it can be predicted that the absence of a given energy term in Eq.~(\ref{eq1}) would result in the absence of the corresponding partial pressure in the equation of state (Eq.~(\ref{eq25})). In particular, by reduction of the long-range RKKY interaction and by leaving only NN magnetic interaction, as well as without energy of electronic subsystem the formalism will be equivalent to that developed in our previous papers (Refs.~\citep{balcerzak_self-consistent_2017, szalowski_thermodynamics_2018}). Such a method could be applicable to magnetic insulators. However, in order to extend the previous formalism to magnetic metals, which has been the purpose of the present paper, the long-range RKKY interaction together with the energy of the electron subsystem have to be taken into account. It should be noted that the remaining energy terms are also important for the full thermodynamic description of the system and cannot be omitted. For instance, the vibrational (Debye) energy is responsible for the correct description of the specific heat. In turn, when the elastic (static) energy is omitted the isothermal compressibility cannot be properly taken into account. 

Thus, as pointed out in the paragraph above, for the full thermodynamic description of the magnetic metals all the energy terms in Eq.~(\ref{eq1}) must exist as a minimum set. Moreover, from the numerical calculations it can be concluded that the electronic subsystem has a great influence on the mechanical and magnetic properties. From one side it is manifested by a weak dependency of the volume deformation on the external pressure, since the Hartree-Fock term leads to a strong compression of the system. On the other hand, the electron concentration, via RKKY interaction, strongly influences the type and properties of magnetic ordering. 

When the magnetic ordering is changed from antiferromagnetic to the ferromagnetic one, as a result of including the NN direct interaction $J_0^d > 0$, the phase transition temperature vs. pressure changes its character from the decreasing to increasing type, respectively (see Fig.~\ref{Fig4}).

\begin{figure}[h]
  \begin{center}
\includegraphics[width=0.7\columnwidth]{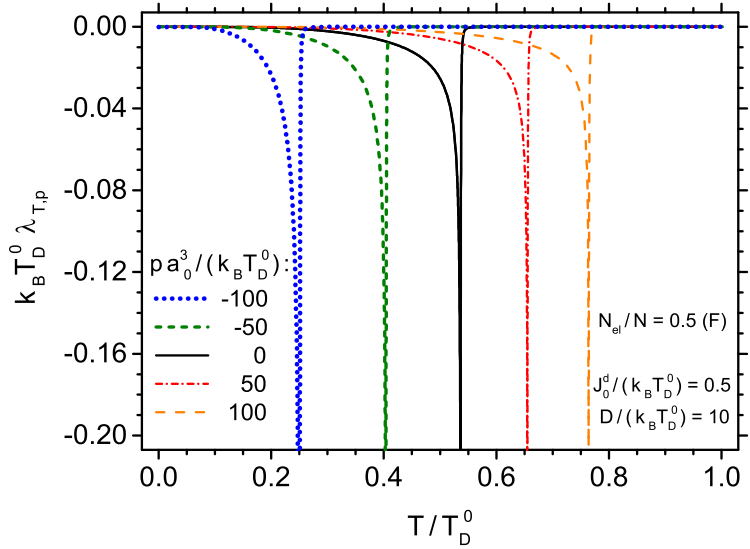}
  \end{center}
   \caption{\label{Fig14}The magnetostriction coefficient, $\lambda_{T,p}$, in dimensionless units, as a function of temperature $\frac{T}{T_{\rm D}^0}$, for ferromagnetic phase, when $\frac{N_{\rm el}}{N}=0.5$. Different curves correspond to various external pressures $\frac{a_0^3}{k_{\rm B}T_{\rm D}^0}p$. The remaining parameters are the same as in Fig.~\ref{Fig13}.}
\end{figure}

\begin{figure}[h]
  \begin{center}
\includegraphics[width=0.7\columnwidth]{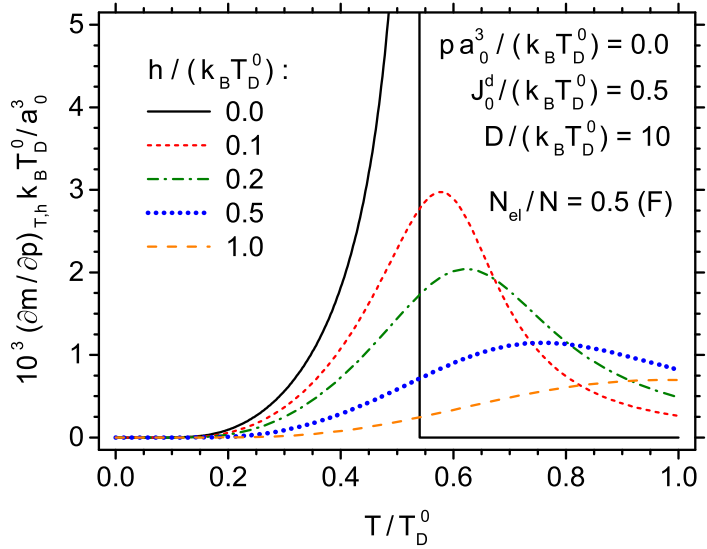}
  \end{center}
   \caption{\label{Fig15}The pressure derivative of magnetization, $\left(\frac{\partial m}{\partial p}\right)_{T,h}$, in dimensionless units, for ferromagnetic phase, when $\frac{N_{\rm el}}{N}=0.5$ and $p=0$. Different curves correspond to various values of the external field $-\frac{g_S\mu_{\mathrm B}H^{z}}{k_{\rm B}T_{\rm D}^0}$. The remaining parameters are the same as in the preceding figure.}
\end{figure}

\begin{figure}[h]
  \begin{center}
\includegraphics[width=0.7\columnwidth]{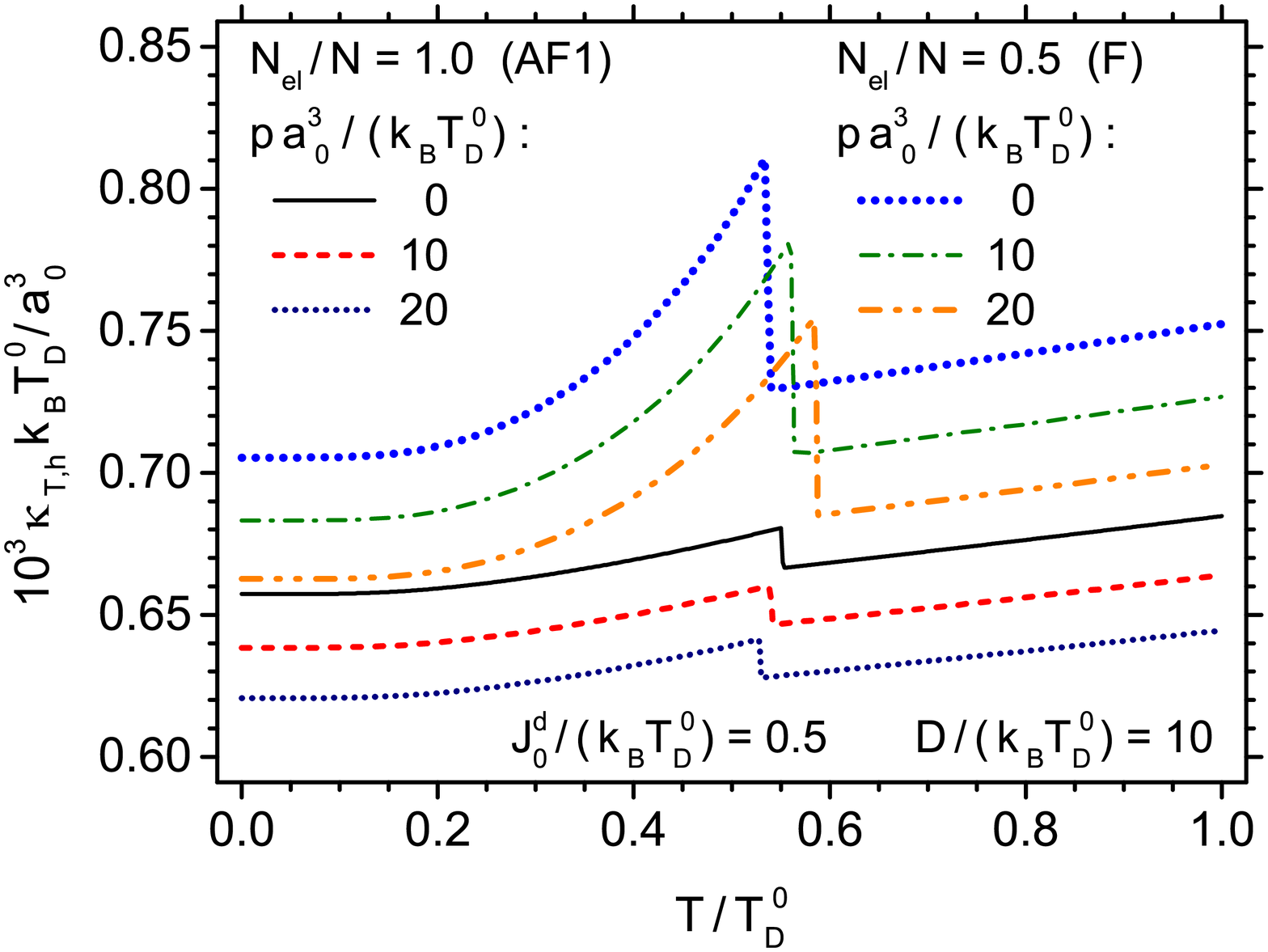}
  \end{center}
   \caption{\label{Fig16}The isothermal compressibility, $\kappa_{T,h}$, in dimensionless units, vs. temperature $\frac{T}{T_{\rm D}^0}$. Different curves correspond to two electron concentrations: $\frac{N_{\rm el}}{N}=1$ (AF1 phase) and $\frac{N_{\rm el}}{N}=0.5$ (ferromagnetic phase), as well as to three different pressures: $\frac{a_0^3}{k_{\rm B}T_{\rm D}^0}p=0$, 10 and 20. The rest of parameters are the same as in the preceding figures.}
\end{figure}

One of the most interesting findings is a decrease of the volume thermal expansion coefficient in  antiferromagnetic phase, when the N\'eel temperature is approached. This invar-like effect may even lead to the negative values of $\alpha_{p,h}$ (see Figs.~\ref{Fig7} and \ref{Fig8}). An explanation of this fact has been given on the basis of the equations of state analysis. We have also checked that such a behaviour represents stable solution, since the total entropy of the system (not shown here) is a monotonically increasing function of temperature. It means that the total specific heat, being a quantity proportional to the entropy derivative over temperature, is positive everywhere, which proves the thermal stability of the system.
It should be mentioned here that negative thermal expansion coefficient has been found experimentally in several materials \cite{bean_magnetic_1962, mary_negative_1996, takenaka_negative_2012}, including frustrated magnets with magnetoelastic couplings \cite{zemen_piezomagnetism_2017}. A decrease of the thermal expansion, leading to the invar effect, is important from the point of view of materials applications and is worth a further study.

The phase transition from antiferromagnetic to the ferromagnetic state, induced by the external field and illustrated in Fig.~\ref{Fig10}, is also worth further study. In particular, the 1st order phase transition occurring in the strong external pressure is an interesting phenomenon. Exact location of the critical field, corresponding to the 1st order transition, can be done on the basis of the Gibbs potential calculations for both ordered phases coexisting in the phase transition point with the same energy.

After analysis of the numerical results it can be stated that, in the presence of RKKY interaction, the influence of magnetoelastic coupling on the thermodynamic properties is significant. The most distinct changes are observed at the phase transition temperature (see, for example, Figs.~\ref{Fig12}, \ref{Fig13} and \ref{Fig14}). All kinds of thermodynamic response functions can be influenced by this coupling, including pure mechanical response, like magnetic susceptibility (Figs.~\ref{Fig11} and \ref{Fig12}) and isothermal compressibility (Fig.~\ref{Fig16}). Examples of a pure thermal response, like specific heat, have been analysed in Ref.\cite{szalowski_thermodynamics_2018} for NN magnetic interactions, and confirm the present conclusion.

The obtained results prove the usefulness of the method, which incorporates energies of various subsystems into a self-consistent thermodynamic description. The method can be further developed for the isotropically deformed systems with higher spins and dilute alloys, both metals and semiconductors as well. In further perspective, an extension of the approach for the crystals with anisotropic volume deformations would be highly desirable.

%\processdelayedfloats

\newpage
\appendix 
\section{Exchange integral $J_k$, effective gyromagnetic factor $g^{\mathrm {eff}}$, and their derivatives over deformation $\varepsilon$}

\subsection{RKKY indirect exchange integral}

The Ruderman-Kittel-Kasuya-Yosida (RKKY) interaction in bulk systems is given by the well known formula \cite{ruderman_indirect_1954,kasuya_theory_1956,yosida_magnetic_1957}
\begin{equation}
\label{A1}
J^{\,\mathrm{RKKY}}_{k}=C\left(k_{\mathrm{F}}a\right)^4\frac {\sin\left(2k_{\mathrm{F}}r_{k}\right)
-2k_{\mathrm{F}}r_{k}\cos\left(2k_{\mathrm{F}}r_{k}\right)}{\left(2k_{\mathrm{F}}r_{k}\right)^{4}}\,
e^{-r_{k}/\lambda}
\end{equation}
In Eq.(\ref{A1}) $a$ is the lattice constant, $r_{k}$ stands for the radius of the $k$-th co-ordination zone and $k_{\mathrm{F}}$ denotes the Fermi wavevector. $C$ is the energy coefficient and the damping parameter $\lambda$ has been introduced by Mattis \cite{mattis_theory_1981} in order to account for the charge carrier localization in disordered systems.
The Fermi wavevector $k_{\mathrm{F}}$ in Eq.(\ref{A1}) is given in the form:
\begin{equation}
\label{A2}
k_{\mathrm{F}}=\left(3\pi^2\frac{N_{\rm el}}{V}\right)^{1/3}=\left(3\pi^2z_0\frac{N_{\rm el}}{N}\right)^{1/3}\frac{1}{a}=k_{\mathrm{F}}^0\frac{1}{\left(1+\varepsilon\right)^{1/3}},
\end{equation}
in which we made use of the equation (\ref{eq4}), whereas $k_{\mathrm{F}}^0$ is defined as:
\begin{equation}
\label{A3}
k_{\mathrm{F}}^0=\left(3\pi^2z_0\frac{N_{\rm el}}{N}\right)^{1/3}\frac{1}{a_0}.
\end{equation}
From the above formulas it is seen that
\begin{equation}
\label{A4}
k_{\mathrm{F}}a=k_{\mathrm{F}}^0a_0=\left(3\pi^2z_0\frac{N_{\rm el}}{N}\right)^{1/3},
\end{equation}
and such a product is volume independent. By the same token, the $k_{\mathrm{F}}r_k$ product can be presented as:
\begin{equation}
\label{A5}
k_{\mathrm{F}}r_k=k_{\mathrm{F}}^0r_{k,0}=\left(3\pi^2z_0\frac{N_{\rm el}}{N}\right)^{1/3}\left(\frac{r_{k,0}}{a_0}\right).
\end{equation}
The energy coefficient $C$ in Eq.(\ref{A1}) is given by the expression
\begin{equation}
\label{A6}
C=\frac{2J^2\left(\frac{V}{N}\right)^{2}m_{\rm el}}{\pi^3\hbar^2a^4},
\end{equation}
in which the energy constant $J$ is the so-called exchange contact potential and $m_{\rm el}$ is the electron mass. It is seen that $C$ is volume dependent and it can be conveniently presented in the form of:
\begin{equation}
\label{A7}
C=C_0\left(\frac{r_{1,0}}{r_0}\right)^2\left(1+\varepsilon\right)^{2/3},
\end{equation}
where $C_0$ is a constant:
\begin{equation}
\label{A8}
C_0=\frac{8J^2 m_{\rm el}}{\pi h^2}\left(\frac{x r_0}{z_0}\right)^2
\end{equation}
whereas $r_{1,0}/r_0$ ratio can be determined from the minimum conditions of the total Gibbs energy for non-deformed structure. Finally, the volume-dependent RKKY interaction can be presented in the form of:
\begin{equation}
\label{A9}
J^{\,\mathrm{RKKY}}_{k}=J^{\,\mathrm{RKKY}}_{k,0}
e^{-x\frac{r_0}{\lambda}\frac{r_{1,0}}{r_0}\frac{r_{k,0}}{a_{0}}\left(1+\varepsilon \right)^{1/3}}\left(1+\varepsilon\right)^{2/3},
\end{equation}
where the volume-independent factor, $J^{\,\mathrm{RKKY}}_{k,0}$, is defined as:
\begin{equation}
\label{A10}
J^{\,\mathrm{RKKY}}_{k,0}=C_0\left(\frac{r_{1,0}}{r_0}\right)^2\left(k_{\mathrm{F}}^0a_0\right)^4\frac {\sin\left(2k_{\mathrm{F}}^0r_{k,0}\right)
-2k_{\mathrm{F}}^0r_{k,0}\cos\left(2k_{\mathrm{F}}^0r_{k,0}\right)}{\left(2k_{\mathrm{F}}^0r_{k,0}\right)^{4}}.
\end{equation}
The derivative of the RKKY exchange integral over $\varepsilon$, which enters the equation (\ref{eq23}), can be now easily found from Eq.(\ref{A9}), namely
\begin{equation}
\label{A11}
\frac{\partial J^{\,\mathrm{RKKY}}_{k}}{\partial \varepsilon}=J^{\,\mathrm{RKKY}}_{k,0}
e^{-x\frac{r_0}{\lambda}\frac{r_{1,0}}{r_0}\frac{r_{k,0}}{a_{0}}\left(1+\varepsilon \right)^{1/3}}\left[-\frac{x}{3}\frac{r_0}{\lambda}\frac{r_{1,0}}{r_0}\frac{r_{k,0}}{a_{0}}+\frac{2}{3}\,\frac{1}{\left(1+\varepsilon \right)^{1/3}}\right]
\end{equation}
(for $k=1,2,\ldots$).

\subsection{NN direct exchange integral}

As far as the NN direct exchange integral, $J^d$, is concerned, it is assumed in the power-law form, analogously to Ref.\cite{balcerzak_self-consistent_2017}:
\begin{equation}
\label{A12}
J^{\rm d}=J^{\rm d}_0 \left( \frac{r_1}{r_{1,0}}\right)^{-n}=J^{\rm d}_0 \left( 1+\varepsilon\right)^{-n/3},
\end{equation} 
where $J^{\rm d}_0$ is the NN direct exchange integral of a non-deformed system (when $T=0$, $p=0$, and $H^z=0$), with $n$ being a constant. 

The derivative of  $J^{\rm d}$ over $\varepsilon$, which also enters the equation (\ref{eq23}) for the magnetic pressure, in addition to $\partial J^{\,\mathrm{RKKY}}_{1}/ \partial \varepsilon$, can now be calculated from Eq.(\ref{A12}) as:
\begin{equation}
\label{A13}
\frac{\partial J^{\rm d}}{\partial \varepsilon}=-\frac{n}{3}J^{\rm d}
\frac{1}{1+\varepsilon}.
\end{equation}

\subsection{Effective gyromagnetic factor}

The effective gyromagnetic factor, $g^{\mathrm {eff}}$, for the RKKY interaction in the presence of the external magnetic field, has been introduced in Ref.~\cite{balcerzak_rkky_2006}. It is given in the form of:
\begin{equation}
\label{A14}
g^{\mathrm {eff}}=g_S\left(1+\frac{g_{\rm el}}{g_S}m_{\rm el} J\frac{\frac{V}{N}k_{\mathrm{F}}}{2 \pi^2 \hbar^2} \right),
\end{equation}
where $g_S$ is the gyromagnetic factor associated with localized spin, and $g_{\rm el} \approx 2$ denotes the gyromagnetic factor of itinerant electron. By substituting $k_{\mathrm{F}}$ from Eq.(\ref{A2}) and $V$ from Eq.(\ref{eq4}), the formula (\ref{A14}) can be presented as:
\begin{equation}
\label{A15}
g^{\mathrm {eff}}=g_S\left(1+\frac{g_{\rm el}}{g_S}\frac{\pi}{4}z_0\frac{C}{J}k_{\mathrm{F}}^0a_0 \right),
\end{equation}
or, equivalently:
\begin{equation}
\label{A16}
g^{\mathrm {eff}}=g_S\left[1+\frac{g_{\rm el}}{g_S}\frac{\pi}{4}z_0\frac{C_0}{J}\left(3\pi^2z_0\frac{N_{\rm el}}{N}\right)^{1/3}\left(\frac{r_{1,0}}{r_0}\right)^2\left(1+\varepsilon\right)^{2/3} \right],
\end{equation}
where $C$ and $C_0$ are given by eqs.(\ref{A7}) and (\ref{A8}), respectively, and $k_{\mathrm{F}}^0a_0$ is expressed by Eq.(\ref{A4}). The last formula  is convenient for calculation of the derivative $\partial g^{\mathrm {eff}}/ \partial \varepsilon$, which appears in Eq.(\ref{eq23}). Then, one obtains:
\begin{equation}
\label{A17}
\frac{\partial g^{\mathrm {eff}}}{\partial \varepsilon}=g_{\rm el}\frac{\pi}{6}z_0\frac{C_0}{J}\left(3\pi^2z_0\frac{N_{\rm el}}{N}\right)^{1/3}\left(\frac{r_{1,0}}{r_0}\right)^2\frac{1}{\left(1+\varepsilon\right)^{1/3}}.
\end{equation}

\section*{Acknowledgments}
This work has been partly supported under grant VEGA 1/033/15.

\newpage
%\bibliography{text}
%\section*{References}

%\bibliographystyle{elsarticle-num}

\end{document}